\documentclass{aa}  
\usepackage{graphicx}
\usepackage{txfonts}
\usepackage[colorlinks=true,
			linkcolor=blue,
			citecolor=blue, 
			filecolor=blue, 
			urlcolor=blue]{hyperref}
\usepackage[switch]{lineno}
\usepackage{txfonts}
\usepackage{natbib}
\usepackage{graphicx, xspace}
\usepackage{multirow}
\usepackage{ amssymb }
\usepackage{subfigure}
\usepackage{epstopdf}
\usepackage{lineno}
\usepackage{cancel}
\usepackage{amsmath}
\pdfmapfile{+txfonts.map}
\usepackage{natbib}
\bibpunct{(}{)}{;}{a}{}{,}

\begin{document} 

\title{MHD study of extreme space weather conditions for exoplanets with Earth-like magnetospheres: On habitability conditions and radio-emission}

\titlerunning{EXTREME SPACE WEATHER IN EARTH-LIKE EXOPLANETS}
\authorrunning{Varela et al.}

   \author{J. Varela\inst{1},
          A. S. Brun\inst{2}, 
          P. Zarka\inst{3},          
          A. Strugarek\inst{2},  
          F. Pantellini\inst{4},          
                   and  
          V. R\'eville\inst{5}   
          }

   \institute{Universidad Carlos III de Madrid, Leganes, 28911 \\
              \email{\href{mailto:jvrodrig@fis.uc3m.es (telf: 0034645770344)}{jvrodrig@fis.uc3m.es}}
          \and  
             Laboratoire AIM, CEA/DRF – CNRS – Univ. Paris Diderot – IRFU/DAp, Paris-Saclay, 91191 Gif-sur-Yvette Cedex, France
          \and
             LESIA \& USN, Observatoire de Paris, CNRS, PSL/SU/UPMC/UPD/UO, Place J. Janssen, 92195 Meudon, France
          \and
             LESIA, Observatoire de Paris, Universit\'e PSL, CNRS, Sorbonne Universit\'e, Universit\'e de Paris, 5 place Jules Janssen, 92195 Meudon, France 
                       \and
            IRAP, Universit\'e Toulouse III—Paul Sabatier, CNRS, CNES, Toulouse, France \\
             }

\date{version of \today}

 
  \abstract
  {
The present study aims at characterizing the habitability conditions of exoplanets with an Earth-like magnetosphere inside the habitable zone of M stars and F stars like $\tau$ Boo, caused by the direct deposition of the stellar wind on the exoplanet surface if the magnetosphere shielding is inefficient. In addition, the radio emission generated by exoplanets with a Earth-like magnetosphere is calculated for different space weather conditions. The study is based on a set of MHD simulations performed by the code PLUTO reproducing the space weather conditions expected for exoplanets orbiting the habitable zone of M stars and F stars type $\tau$ Boo. Exoplanets hosted by M stars at $0.2$ au are protected from the stellar wind during regular and CME-like space weather conditions if the star rotation period is slower than $3$ days, that is to say, faster rotators generate stellar winds and interplanetary magnetic fields large enough to endanger the exoplanet habitability. Exoplanets hosted by a F stars type $\tau$ Boo at $\geq 2.5$ au are protected during regular space weather conditions, but a stronger magnetic field compared to the Earth is mandatory if the exoplanet is close to the inner edge of the star habitable zone ($2.5$ au) to shield the exoplanet surface during CME-like space weather conditions. The range of radio emission values calculated in the simulations are consistent with the scaling proposed by \citet{Zarka8} during regular and common CME-like space weather conditions. If the radio telescopes measure a relative low radio emission signal with small variability from an exoplanet, that may indicate favorable exoplanet habitability conditions with respect to the space weather states considered and the intrinsic magnetic field of the exoplanet. The radio emission power calculated for exoplanets with an Earth-like magnetosphere inside the star habitable zone  is in the range of $3 \cdot 10^{7}$ to $2 \cdot 10^{10}$ W if the space weather conditions lead to SW dynamic pressures between $1.5$ to $100$ nPa and IMF intensities between $50$ - $250$ nT, and is below the sensitivity threshold of present radio telescopes at parsec distances.
    }


\maketitle


\section{Introduction}

The space weather effects on the Earth magnetosphere were extensively studied in the last years \citep{Poppe,Gonzalez,Varela7}, particularly during extreme events such as intense coronal mass ejections (CME) \citep{Low,Howard} leading to major perturbations in the Earth magnetosphere structures \citep{Wang4,Lugaz,Wu}. 

The CMEs are solar eruptions produced in the corona due to magnetic reconnections, expelling fast charged particles and a magnetic cloud \citep{Neugebauer,Cane2,Regnault}. Extreme space weather events are not exclusive of the Sun or solar-like stars \citep{Leitzinger}, CMEs were also observed in M, K and F type stars \citep{Khodachenko,Lammer}.

The space weather at the orbit of the Earth and exoplanets depends on the stellar wind (SW) and interplanetary magnetic field (IMF) generated by the host star \citep{Strugarek,Garraffo} at their orbital location as well as the conducting and magnetic properties of the local environment. For the case of the Earth, the intrinsic magnetic field is strong enough to avoid the direct precipitation of the SW on the surface even during the largest CMEs observed \citep{Salman,Kilpua,Hapgood}. Extreme space weather conditions occur if the SW dynamic pressures in the range of the $10$ to $100$ nPa and IMF intensity between $100$ and $300$ nT.

The space weather in the orbit of exoplanets cannot be compared to the case of the Earth if the host star has characteristics different from the Sun (star type, age, metallicity, ...). If the SW dynamic pressure and IMF intensity generated by the star are large, favorable exoplanet habitability state requires an intrinsic magnetic field strong enough to avoid the direct precipitation of the SW on the exoplanet surface \citep{Gallet,Linsky,Airapetian}. Otherwise, if the protection of the magnetic field is deficient, the exoplanet habitability can be hampered by the effect of the SW as well as the depletion of the atmosphere, especially volatile components such as the water molecules \cite{Lundin,Moore,Jakosky}. It should be noted that other important factors for the habitability as EUV, X ray and cosmic rays fluxes towards the exoplanet surface are not included in the analysis as such effects are beyond the scope of the present study. Nevertheless,  the eventual direct precipitation of the SW must be understood as an important constraint for the habitability of planets.

Exoplanet habitability could be constrained for exoplanet without an intrinsic magnetic field, although the detection and characterization of exoplanet magnetospheres is a challenging topic. It is known from the interaction of the SW with the planets of the solar system that intrinsic magnetic fields are emitters of cyclotron MASER emission at radio wavelengths \citep{Kaiser,Zarka5,Lamy}, generated by energetic electrons accelerated in the reconnection region between IMF and the planet magnetic field, flowing towards the planet surface along the magnetic field lines \citep{Wu5}. A fraction of the electrons energy is transformed into cyclotron radio emission \citep{Zarka5} escaping from the magnetosphere. Such radio emission is detected by ground-based radio telescopes, for example the Nançay decameter array \citep{Lamy}, NenuFAR \citep{Zarka11} and Low Frequency Array (LOFAR) \citep{Haarlem} between others. Likewise, the radio emission detected from an exoplanet magnetosphere could provide information of the exoplanet intrinsic magnetic field \citep{Hess}. Unfortunately, the detection capability of present radio telescopes barely distinguish the radio emission from exoplanets. Recent LOFAR and the Australian Telescope Compact Array (ATCA) measurements tentatively achieved the detection of radio emission from exoplanet systems \citep{Turner3,Torres}. In addition, radio emission from the red draft GJ $1151$ was measured, potentially originated in the magnetic interaction with a exoplanet with approximately the size of the Earth \citep{Vedantham,Pope,Perger}. Next generation of radio telescopes may be able to detect exoplanet radio emissions at a distances of 20 parsec \citep{Carilli,Nan,Ricci2,Zarka11}, for example the Square Kilometre Array (SKA) \citep{Zarka10}, depending on the space weather conditions generated by the host star and the properties of the exoplanet magnetic field.

This study is the continuation of a research activity dedicated to analyze numerically the interaction of the stellar wind with planetary magnetospheres, particularly the radio emission generation with respect to the space weather conditions and the properties of the planet intrinsic magnetic field. First, the radio emission from the Hermean magnetosphere was analyzed in \citet{Varela5}, showing the important role of the IMF intensity, IMF orientation and SW dynamic pressure on the radio emission generated. Then, \citet{Varela6} was dedicated to study the radio emission from exoplanets with different intrinsic magnetic field configurations, identifying a critical dependency between magnetosphere topology and radio emission. Next, \citet{Varela7} analyzed the effect of extreme space weather conditions on the Earth magnetosphere. The aim of the present study is to analyze the effect of the space weather conditions on the magnetosphere of exoplanets orbiting the habitable zone of M and F stars. In addition, the radio emission generated from the exoplanet magnetosphere is estimated. The analysis consist in a set of MHD simulations assuming the exoplanet magnetic field is identical to the Earth magnetic field, reproducing the space weather conditions inside the habitable zone of M and F stars.

This paper is structured as follows. Section 2 presents the description of the numerical model. Section 3 introduces the analysis of the space weather effects on the magnetosphere of exoplanet orbiting the habitable zone of M and F stars. Section 4 presents the characterization of the radio emission generated by exoplanets with an Earth-like magnetosphere during extreme space weather conditions. Section 5 discusses and concludes the analysis results.

\section{Numerical model}

This study is performed using the ideal MHD version of the open-source code PLUTO in spherical coordinates. The model calculates the evolution of a single-fluid polytropic plasma in the nonresistive and inviscid limit \citep{Mignone}. A detailed description of the model equations, boundary conditions and upper ionosphere model can be found in \citep{Varela7}.

The interaction of the SW with planetary magnetospheres can be studied using different numerical models; present study uses a single fluid MHD code \citep{2008Icar..195....1K,2015JGRA..120.4763J,Varela,Strugarek2,Strugarek}. The validity of MHD code results were checked by comparing the simulation results with ground-based magnetometers and spacecraft measurements \citep{Watanabe,Raeder2,Wang6,Facsko}. The study was performed using the single-fluid MHD code PLUTO in spherical 3D coordinates \citep{Mignone}. The model was applied successfully to study the global structures of the Hermean magnetosphere \citep{Varela,Varela2,Varela3,Varela4,Varela5},the radio emission from exoplanets \cite{Varela6} and the effect of extreme space weather conditions on the Earth magnetosphere \citep{Varela7}.

The simulations use a grid of 128 radial points, 48 in the polar angle $\theta$ and 96 in the azimuthal angle $\phi$, equidistant in the radial direction. The simulation domain is confined between two concentric shells around the exoplanet, with the inner boundary $R_{in} = 2R_{ex}$ ($R_{ex}$ the exoplanet radius) and the outer boundary $R_{out} = 30R_{ex}$. The upper ionosphere model extends between the inner boundary and $R = 2.5R_{ex}$.

The exoplanet magnetic field is rotated $90^{o}$ in the YZ plane with respect to the grid poles with the aim of avoiding numerical issues (no special treatment was included for the singularity at the magnetic poles). The exoplanet magnetosphere is identical to the Earth magnetosphere, thus the tilt of the Earth rotation axis is also included ($23^{o}$ with respect to the ecliptic plane).

The simulation frame assumed is: z-axis is provided by the planetary magnetic axis pointing to the magnetic north pole, star-planet line is located in the XZ plane with $x_{star} > 0$ (solar magnetic coordinates) and the y-axis completes the right handed system.

The response of the exoplanet magnetosphere for different SW dynamic pressure ($P_{d}$), IMF intensity ($|B|_{IMF}$) and orientation is calculated based on the data regression obtained by the set of simulations performed in \citet{Varela7} (see Table 5). The SW dynamic pressure is defined as $P_{d} = m_{p} n_{sw}  v_{sw}^{2}/2$, with $m_{p}$ the proton mass, $n_{sw}$ the SW density and $v_{sw}$ the SW velocity.

The effect of different IMF orientations are included in the analysis: Exoplanet-star and star-exoplanet (also called radial IMF configurations), southward, northward and ecliptic clockwise. Exoplanet-star and star-exoplanet configurations indicate an IMF parallel to the SW velocity vector. Southward and northward IMF orientations show an IMF perpendicular to the SW velocity vector in the XZ plane.

\section{Magnetopause standoff distance for exoplanets with an Earth-like magnetic field}

This section is dedicated to calculate the magnetopause standoff distance of exoplanets with an Earth-like magnetic field exposed to different space weather conditions. A detailed description of the standoff distance calculation in the simulations is shown in the appendix.  The analysis includes regular and CME-like space weather conditions expected for exoplanet orbiting inside the habitable zone of M and F stars. Consequently, the study provides a first order assessment of the exoplanet habitability with respect to the SW direct deposition on the exoplanet surface. The analysis is performed assuming exoplanets with an Earth-like magnetic field because no observational data exists regarding the properties of exoplanets magnetosphere. Nevertheless, the different IMF orientations tested are equivalent to exoplanets with different tilt angles.

The space weather conditions inside the stellar habitable zone change with the star characteristics \citep{Kasting,Tarter,Kopparapu,Johnstone,Cuntz,Airapetian}. The habitable zone for main sequence F stars ($1.1$ – $1.5 M_{Sun}$) is located between $2.5$ - $5$ au \citep{Sato}, G stars ($1.1$ - $0.9 M_{Sun}$) between $0.84$ – $1.68$ au \citep{Kopparapu2}, K stars ($0.9$ – $0.5 M_{Sun}$) between $0.21$ – $1.27$ au \citep{Cuntz} and M stars ($< 0.5 M_{Sun}$) between $0.03$ – $0.25$ au \citep{Shields}. In the following, the habitability conditions imposed by the star in exoplanets at different orbits inside the habitable zone of M and F stars are studied.

The habitability conditions obtained in the simulations are defined with respect to the magnetopause standoff distance above the exoplanet surface. If the normalized standoff distance is $R_{mp} / R_{ex} = 1$ ($R_{mp}$ is the exoplanet magnetopause standoff distance) there is a direct precipitation of the SW towards the exoplanet surface. This is the same criteria used in \citet{Varela7} (equations $5$ and $6$).

\subsection{Exoplanet hosted by M stars}

M type stars habitability conditions are an open issue because exoplanets inside the habitable zone are likely to be tidally locked \citep{Griemeier3,Griemeier4} and exposed to a strong radiation from the host star \citep{Scalo} as well as persistent CME events \citep{Khodachenko,Lammer}. Nevertheless, recent studies indicate tidal locking may constrain but not preclude the habitability conditions of exoplanets\citep{Yang,Hu2,Leconte,Barnes}. Previous studies also assessed the space weather conditions in the orbit of exoplanets inside the habitable zone of M stars \citep{Odstrcil2,Odstrcil,Vidotto9}. Table\ref{1} shows the density, velocity and dynamic pressure of the SW generated by a M star at different orbits following \citet{Johnstone2} SW model for regular and CME-like space weather conditions. The CME-like space weather conditions are guess educated values assuming $20$ times the SW density and $2.5$ times the SW velocity of the regular space weather conditions. Such parameters are typical for CME conditions for the Sun.

\begin{table}
\centering
\begin{tabular}{c | c c c}
 &  & Regular SW & \\ \hline
AU & $n_{sw}$ & $|v_{sw}|$ & $P_{d}$\\
 & (cm$^{-3}$) & (km/s) & (nPa) \\ \hline
$0.05$ & $2000$ & $540$ & $488$ \\
$0.1$ & $500$ & $650$ & $177$ \\
$0.2$ & $90$ & $700$ & $37$ \\ \hline
 &  & CME-like SW & \\ \hline
AU & $n$ & $|v|$ & $P_{d}$\\
 & ($10^{3}$ cm$^{-3}$) & (km/s) & ($10^{3}$ nPa) \\ \hline 
$0.05$ & $40$ & $1350$ & $61$  \\
$0.1$ & $10$ & $1650$ & $23$ \\
$0.2$ & $1.8$ & $1750$ & $4.6$ \\ 
\end{tabular}
\caption{Exoplanet orbit inside the habitable zone of M stars (first column). SW density (second column), velocity (third column) and dynamic pressure (fourth column) for regular and CME-like space weather conditions.}
\label{1}
\end{table}

Figure \ref{1} shows the exoplanet habitability constrain imposed by the space weather conditions inside the habitable zone of a M star. The graphs indicate the critical IMF intensity and SW dynamic pressure required for the direct SW precipitation towards the exoplanet surface in the equatorial region (for different IMF orientations), that is to say, the space weather conditions leading to a normalized exoplanet magnetopause standoff distance of $R_{mp}/R_{ex}=1$. It should be noted that the graphs show the data regression obtained by the simulation performed in \citet{Varela7}, dedicated to calculate the Earth magnetopause standoff distance for different values of the SW dynamic pressure, IMF intensities and IMF orientations. The range of SW dynamic pressure and IMF intensity values included in the study correspond to regular (panel a) and CME-like (panel b) space weather conditions. The horizontal dashed lines indicate the SW dynamic pressure at the orbit of an exoplanet at $0.05$ au (red), $0.1$ au (orange) and $0.2$ au (blue) from the host star based on \citet{Johnstone2} SW model, providing a reference value of the critical IMF intensity required for the direct SW precipitation onto the exoplanet surface for different IMF orientations based on the pressure balance (see appendix).

\begin{figure}[h]
\centering
\resizebox{\hsize}{!}{\includegraphics[width=\columnwidth]{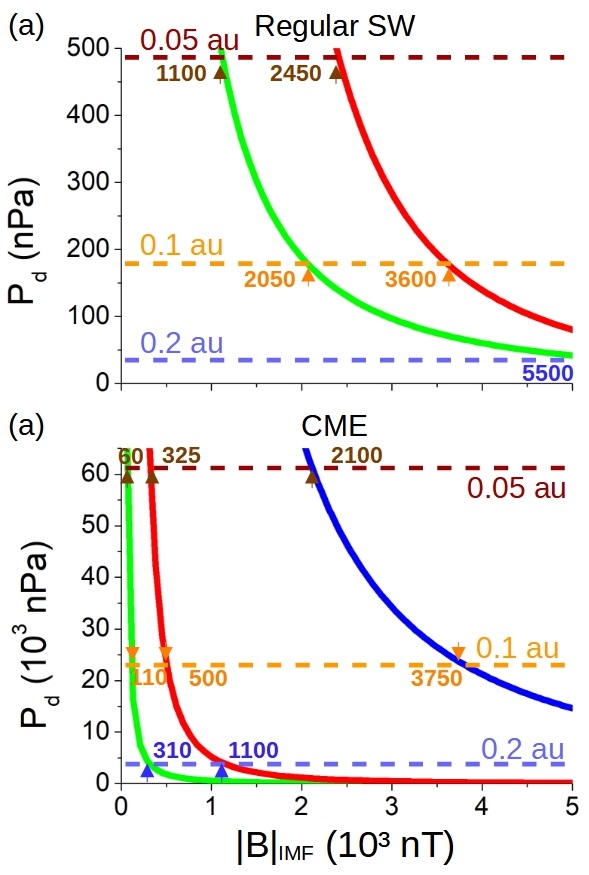}}
\caption{Critical IMF intensity and SW dynamic pressure required for the direct precipitation of the SW towards the exoplanet surface for (a) regular and (b) CME-like space weather conditions. IMF orientation: Exoplanet-star (red line), southward (green line) and northward (blue line). The horizontal dashed lines indicate the SW dynamic pressure at different exoplanet orbits: $0.05$ au (red), $0.1$ au (orange) and $0.2$ au (blue). The critical IMF intensity is indicated for each IMF orientation.}
\label{1}
\end{figure}

During regular space weather conditions, panel a, the critical IMF intensity for an exoplanet at $0.2$ au is $|B|_{IMF} > 5000$ nT , $\approx 2050$ nT at $0.1$ au and $\approx 1100$ nT at $0.05$ au if the IMF is southward. The southward IMF is highlighted along the article because it is the IMF orientation leading to the lowest magnetopause standoff distance (maximum reconnection) for a fixed IMF intensity. Consequently, the magnetic field generated by M stars must be very large to threaten the exoplanet habitability. Nevertheless, the magnetic field of young and fast rotating M stars can overcome such IMF intensity thresholds \citep{Shulyak,Shulyak2} reaching values up to $4$ kG. The IMF intensity threshold during a CME largely decreases compared to regular space weather conditions, panel b. If the exoplanet orbit is at $0.2$ au, the critical $|B|_{IMF} \approx 310$ nT for a southward IMF and $\approx 1100$ nT for a star-exoplanet IMF. If the exoplanet is at $0.1$ au, $|B|_{IMF} \approx 110$ nT for a southward IMF, $\approx 500$ nT for a star-exoplanet IMF and $\approx 3750$ nT for a northward IMF. If the exoplanet is at $0.05$ au, $|B|_{IMF} \approx 60$ nT for a southward IMF, $\approx 325$ nT for a star-exoplanet IMF and $\approx 2100$ nT for a northward IMF. That is to say, exoplanets at $0.2$ au are efficiently protected during CME space weather conditions if the intensity of the magnetic field generated by the M star is not strong enough to exceed $310$ nT. On the other hand, exoplanets at $\le 0.1$ au are exposed to the direct SW precipitation during CMEs if the IMF intensity exceeds $110$ nT. In summary, exoplanets at $0.2$ au should be protected from the direct precipitation of the SW by an Earth-like magnetic field, thus the exoplanets is habitable with respect to the SW shielding. It should be noted that present study conclusions are consistent with respect to configuration subsets analyzed by other authors \citep{Garraffo,Garraffo2}.

As it was mentioned in the previous paragraph, the space weather conditions change with the rotation rate of the star, because the magnetic activity and the properties of the SW generated by the star change \citep{Suzuki}. The SW velocity during regular space weather conditions is $2$ times larger if the star rotation is $4$ times faster, although the SW density and temperature is weakly affected \citep{Shoda}. In addition, faster rotators have a stronger magnetic activity, because the large-scale surface magnetic field ($B_{surf,*}$) dependency with the Rossby number ($R_{o}$) is $B_{surf,*} \propto R_{o}^{-1.3}$ \citep{See,Brun}. Thus the IMF intensity at the exoplanet orbit is higher as well as the CME frequency and intensity \citep{Shulyak,Shulyak2}. Consequently, if the effect of the M star rotation period is included in the analysis, the threshold of the IMF intensity and SW dynamic pressure for the direct precipitation of the SW toward the exoplanet surface changes. Table \ref{2} indicates the SW density and velocity in the orbit of an exoplanet at $0.05$, $0.1$ and $0.2$ au from the host M star for different rotation periods ($P_{rot}$) for the star during regular space weather conditions (data derived from \citet{Shoda} simulations). The SW density has a weak dependency with the star rotation but the SW velocity and IMF intensity increases with the star rotation. The range of M star rotation periods analyzed include the majority of the $795$ M stars identified by Kepler mission as a sub-sample of the $12000$ main sequence stars identified \citep{Nielsen}. Nevertheless, recent surveys of M star identified an important population of slow M stars rotators, showing rotation periods between $30$ to $120$ days \citep{Newton,Popinchalk}.

\begin{table}
\centering
\begin{tabular}{c | c c c c c}
AU & $P_{rot}$ & $n_{sw}$ & $|v_{sw}|$ & $P_{d}$ & $|B|_{IMF}$\\
 & (days) & (cm$^{-3}$) & (km/s) & (nPa) & ($10^{3}$ nT)\\ \hline
$0.05$ & $24$ & $4500$ & $280$ & $295$ & $2.16$ \\
$0.05$ & $12$ & $4500$ & $360$ & $488$ & $17.7$ \\
$0.05$ & $6$ & $4500$ & $400$ & $602$ & $25.9$ \\
$0.05$ & $3$ & $4500$ & $450$ & $762$ & $30.3$ \\ \hline
$0.1$ & $24$ & $900$ & $350$ & $92.2$ & $0.54$ \\
$0.1$ & $12$ & $900$ & $440$ & $146$ & $4.43$ \\
$0.1$ & $6$ & $900$ & $510$ & $196$ & $6.46$ \\
$0.1$ & $3$ & $900$ & $620$ & $289$ & $7.57$ \\ \hline
$0.2$ & $24$ & $240$ & $410$ & $33.7$ & $0.31$ \\
$0.2$ & $12$ & $240$ & $500$ & $50.2$ & $1.11$ \\
$0.2$ & $6$ & $240$ & $590$ & $69.9$ & $1.62$ \\
$0.2$ & $3$ & $240$ & $800$ & $128$ & $1.89$ \\
\end{tabular}
\caption{Exoplanet orbit inside the habitable zone of M stars (first column). Star rotation period (second column). SW density (third column), velocity (forth column) and dynamic pressure (fifth column). IMF intensity (sixth column).}
\label{2}
\end{table}

Figure \ref{2} indicates the IMF intensity and SW dynamic pressure threshold with respect to the M star rotation rate for regular space weather conditions.

\begin{figure}[h]
\centering
\resizebox{\hsize}{!}{\includegraphics[width=\columnwidth]{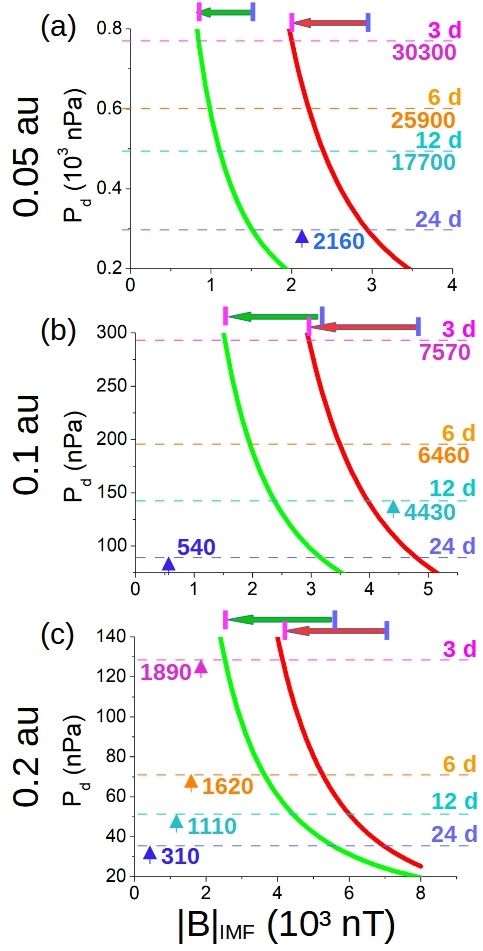}}
\caption{Critical IMF intensity and dynamic pressure required for the direct precipitation of the SW considering different M star rotation periods and exoplanets located at (a) $0.05$ au, (b) $0.1$ au and (c) $0.2$ au orbits. IMF orientation: Exoplanet-star (red line) and southward (green line). The horizontal dashed lines indicate the SW dynamic pressure for M stars with rotation periods: $24$ days (blue), $12$ days (light cyan), $6$ days (orange) and $3$ days (pink). The bold colored arrows show the decrease of the critical IMF intensity required for the direct SW deposition if the M star rotation increases from $24$ to $3$ days. The green (red) color of the bold horizontal arrow indicates a southward (exoplanet-star) IMF orientation. The critical IMF intensity following \citet{Shoda} simulations is indicated for each star rotation rate.}
\label{2}
\end{figure}

The model shows a large decrease of the IMF intensity threshold if the M star rotation period decreases given a SW dynamic pressure. $\Delta |B|_{IMF}$ is indicated by the bold arrows in the top of the graph for each IMF orientation between the cases of star with rotation rates of $24$ and $3$ days. For an exoplanet at $0.05$ au, the IMF intensity threshold decreases from $1500$ nT to $850$ nT reducing the star rotation period from $24$ to $3$ days if the IMF is southward, as well as from $3000$ nT to $2000$ nT if the IMF is in the exoplanet-star orientation. Regarding an exoplanet orbit at $0.1$ au, the IMF intensity threshold decreases from $3250$ nT to $1500$ nT for a southward IMF, as well as from $4750$ nT to $3000$ nT for an exoplanet-star IMF. If the exoplanet orbit is located at $0.2$ au, the IMF intensity threshold decreases from $5550$ nT to $2600$ nT for a southward IMF and from $7000$ nT to $4250$ nT for an exoplanet-star IMF. The IMF intensity threshold obtained can be compared with the magnetic field generated by M stars at different orbits following \citet{Shoda} simulations (last column of table \ref{2}). At $0.05$ au, the IMF intensity is above the threshold for a Southward IMF orientation if the star rotation period is shorter than $24$ days, and below the threshold for an exoplanet-star IMF if the rotation period is $24$ days or larger. That is to say, favorable habitability conditions with respect to SW of an exoplanet at $0.05$ au require an intrinsic magnetic field stronger than Earth´s if the rotation rate of the M star is $24$ days or smaller. At $0.1$ au, the IMF intensity is above the threshold for Southward and exoplanet-star IMF orientation and the rotation rate is $12$ days or faster. Thus, exoplanets at $0.1$ au require a magnetic field stronger than the Earth if the host M star rotation rate is smaller than $12$ days. If the exoplanet is at $0.2$ au, the IMF intensity is below the threshold for all IMF orientations if the star rotation rate is $3$ days or slower, so an Earth-like magnetic field can efficiently shield the exoplanet surface.

Summarizing, exoplanets with an Earth-like magnetic field hosted by a M star and located at $0.2$ au are shielded from the SW during regular and CME-like space weather conditions. In addition, such protection holds for M stars with rotation periods as fast as $3$ days during regular SW space weather conditions. Nevertheless, fast rotating M stars with strong and recurrent CME-like events can restrict the exoplanet habitability conditions. On the other hand, exoplanets at $0.1$ au are shielded from regular and CME-like space weather conditions only if the M stars rotation period is $12$ days or larger. Finally, exoplanets at $0.05$ are vulnerable during CME-like events even for M stars with the a rotation period of $24$ days, thus exoplanet habitability requires a magnetic field stronger with respect to the Earth. Nevertheless, exoplanet at $0.05$ au hosted by slower rotators with $P_{rot} > 24$ days are protected during standard and CME-like events by an Earth-like magnetic field if the IMF intensity is lower than $1000$ nT for a southward IMF.

\subsection{Exoplanet hosted by F stars type $\tau$ Boo}

Space weather conditions in F stars were analyzed in previous studies, particularly for $\tau$ Boo type $F7V$, concluding the SW may have a density $135$ times larger with respect to the SW generated by the Sun, as well as a velocity around $300$ km/s \citep{Vidotto6}. Table\ref{3} shows guess educated values of the space weather conditions in the orbit of an exoplanet hosted by a F star similar to $\tau$ Boo near the bottom and upper range of the habitable zone. The SW density during regular space weather conditions is assumed $100$ times the SW density generated by the Sun at $2.5$ and $5$ au. The velocity is the same with respect to \citep{Vidotto6}, $300$ km/s at $2.5$ au. In addition, an extrapolation is assumed to characterize the space weather conditions during CMEs, selecting a SW density $20$ times larger and a velocity $5$ times higher with respect to the regular space weather conditions.

\begin{table}
\centering
\begin{tabular}{c | c c c}
 &  & Regular SW & \\ \hline
AU & $n_{sw}$ & $|v_{sw}|$ & $P_{d}$\\
 & (cm$^{-3}$) & (km/s) & (nPa) \\ \hline
$2.5$ & $50$ & $300$ & $3.8$  \\
$5.0$ & $20$ & $310$ & $1.6$ \\ \hline
 &  & CME-like SW & \\ \hline
AU & $n$ & $|v_{sw}|$ & $P_{d}$\\
 & ($10^{3}$ cm$^{-3}$) & ($10^{3}$ km/s) & ($10^{3}$ nPa) \\ \hline
$2.5$ & $1.0$ & $1.5$ & $1.88$  \\
$5.0$ & $0.4$ & $1.55$ & $0.8$ \\

\end{tabular}
\caption{Exoplanet orbit inside the habitable zone of F star type $\tau$ Boo (first column). SW density (second column), velocity (third column) and dynamic pressure (fourth column) for regular and CME-like space weather conditions.}
\label{3}
\end{table}

Figure \ref{3} indicates the critical IMF intensity and SW dynamic pressure required for the direct SW precipitation towards an exoplanet hosted by a F star type $\tau$ Boo inside the habitable zone during CME-like space weather conditions. The same analysis for regular space weather conditions is not included because the IMF intensity and SW dynamic pressure are well below the threshold required for the direct SW precipitation, that is to say, the exoplanets at $2.5 - 5.0$ au are shielded during regular space weather conditions.  

\begin{figure}[h]
\centering
\resizebox{\hsize}{!}{\includegraphics[width=\columnwidth]{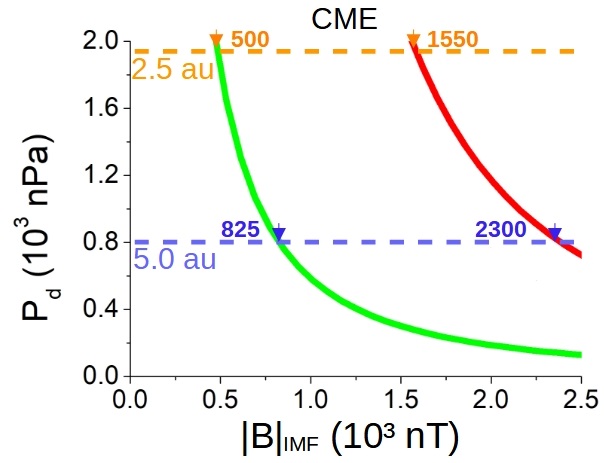}}
\caption{Critical IMF intensity and SW dynamic pressure required for the direct precipitation of the SW towards the exoplanet surface during CME-like space weather conditions. IMF orientation: Exoplanet-star (red line) and southward (green line). The horizontal dashed lines indicate the SW dynamic pressure at different exoplanet orbits: $2.5$ au (orange) and $5.0$ au (blue). The critical IMF intensity is indicated for each IMF orientation.}
\label{3}
\end{figure}

Exoplanets located at $5$ au show an IMF intensity threshold of $|B|_{IMF} \approx 825$ nT for a southward IMF and $|B|_{IMF} \approx 2300$ nT for an exoplanet-star IMF. Regarding exoplanets at $2.5$ au, the IMF intensity threshold is $|B|_{IMF} \approx 500$ nT for a southward IMF and $|B|_{IMF} \approx 1550$ nT for an exoplanet-star IMF. It must be noted the magnetic activity of $\tau$ Boo is larger with respect to the Sun, showing a shorter magnetic cycle of $2$ years \citep{Fares,Fares2}. It is known that F stars have a slower decrease of the rotation rate along the main sequence, leading to a stronger magnetic field compared to G stars \citep{Saffe,Mathur} with the exception of low mass stars populations ($<0.9 M_{Sun}$) that maintain rapid rotation for much longer than solar-mass stars \citep{Matt}. Consequently, the effect of the CME on exoplanets orbiting inside the habitable zone of F star, particular $\tau$ Boo, can ¡increase the exoplanet habitability conditions if the frequency of these extreme space weather events is high.

Next step of the analysis is to include the effect of stellar rotation. The F star rotation period is lower with respect to less massive stars such as G, K and M stars. The lower bound is around $2$ days for $F0$ stars increasing to $10$ days for $F9$ stars  \citep{Nielsen}. Table \ref{4} indicates guess educated values of the SW dynamic pressure and IMF intensity at different exoplanet orbits for different F star rotation periods during CME space weather conditions. The values of the IMF intensity are extrapolated from observational data of F stars magnetic field magnitude \citep{Bailey,Mathur,Marsden,See,Seach} and modeling results \citep{Brun}. We assume the SW velocity increases with the star rotation although the SW density and temperature is constant, extrapolating \citet{Shoda} results.

\begin{table}
\centering
\begin{tabular}{c | c c c c c}
AU & $P_{rot}$ & $n_{sw}$ & $|v_{sw}|$ & $P_{d}$ & $|B|_{IMF}$\\
 & (days) & ($10^{3}$ cm$^{-3}$) & ($10^{3}$ km/s) & ($10^{3}$ nPa) & ($10^{3}$ nT)\\ \hline
$2.5$ & $2$ & $1.0$ & $1.7$ & $2.4$ & $3$ \\
$2.5$ & $5$ & $1.0$ & $1.3$ & $1.4$ & $1.5$ \\
$2.5$ & $7.5$ & $1.0$ & $1.15$ & $1.1$ & $1$ \\
$2.5$ & $10$ & $1.0$ & $1.0$ & $0.8$ & $0.5$ \\ \hline
$5.0$ & $2$ & $0.4$ & $1.75$ & $1.0$ & $0.75$ \\
$5.0$ & $5$ & $0.4$ & $1.35$ & $0.6$ & $0.4$ \\
$5.0$ & $7.5$ & $0.4$ & $1.2$ & $0.5$ & $0.25$ \\
$5.0$ & $10$ & $0.4$ & $1.05$ & $0.4$ & $0.1$ \\
\end{tabular}
\caption{Exoplanet orbit inside the habitable zone of F star (first column). Star rotation period (second column). SW density (third column), velocity (forth column) and dynamic pressure (fifth column). IMF intensity (sixth column).}
\label{4}
\end{table}

Figure \ref{4} indicates the IMF intensity and SW dynamic pressure threshold with respect to the F star rotation rate for CME-like space weather conditions.

\begin{figure}[h]
\centering
\resizebox{\hsize}{!}{\includegraphics[width=\columnwidth]{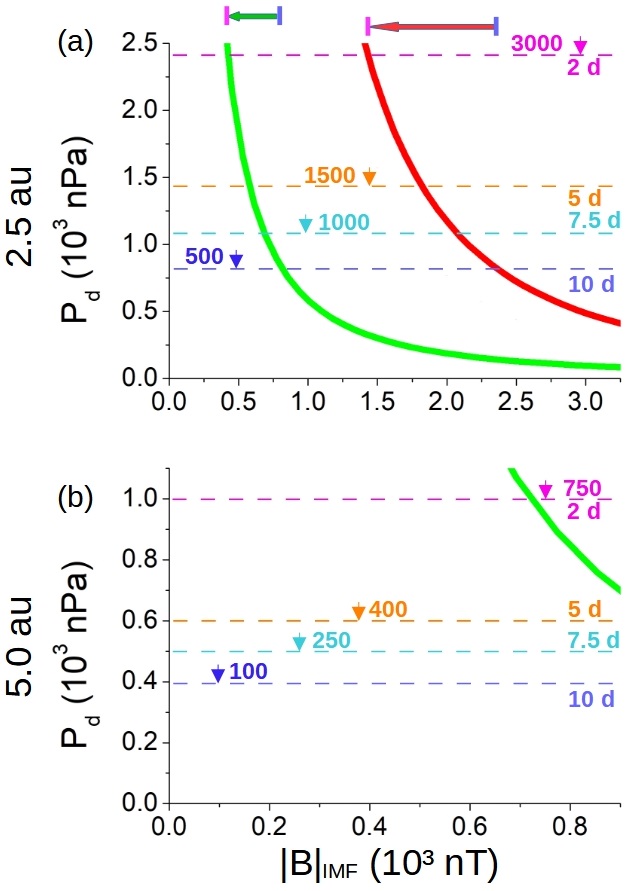}}
\caption{Critical IMF intensity and dynamic pressure required for the direct precipitation of the SW considering different F star rotation periods and exoplanets located at $2.5$ au (a) and $5.0$ au (b) orbits. IMF orientation: Exoplanet-star (red line) and southward (green line). The horizontal dashed lines indicate the SW dynamic pressure for F stars with rotation periods: $10$ days (blue), $7.5$ days (light cyan), $5$ days (orange) and $2$ days (pink). The bold colored arrows show the decrease of the critical IMF intensity required for the direct SW deposition if the F star rotation increases from $10$ to $2$ days. The green (red) color of the bold arrow indicates a southward (exoplanet-star) IMF orientation. The tentative critical IMF intensity is indicated for each star rotation rate.}
\label{4}
\end{figure}

The simulations indicate the habitability of exoplanets at $2.5$ au from the host F star is conditioned by the SW if the star rotation period is shorter than $10$ days. The exoplanet surface is protected if the star rotation period is $10$ days or above, showing an IMF intensity of $500$ nT that is smaller compared to the IMF intensity required for the direct SW precipitation . For a stellar rotation of $7.5$ or $5$ days, direct SW precipitation exists during a southward IMF with $675$ and $575$ nT, respectively, smaller than the IMF intensity during CMEs. The IMF threshold for the direct SW precipitation is also largely exceeded if the star rotation is $2$ days for an IMF oriented in the Southward or Exoplanet-star directions. Consequently, exoplanets at $2.5$ au requires an intrinsic magnetic field intensity stronger with respect to the Earth if the star rotation period is smaller than $10$ days. On the other hand, the simulations show that exoplanets with orbits at $5.0$ au are protected during CME-like space weather conditions if the star rotation period is above $2$ days. In the case of the rotation period is $2$ days the IMF intensity threshold is similar to the IMF intensity during CMEs (around $25$ nT smaller).

In summary, regular space weather conditions does not impact the habitability of exoplanets in the habitable zone of F stars type $\tau$ Boo. On the other hand, persistent and strong CME events can largely influence the habitability of exoplanets nearby the inner boundary of the habitable zone, thus a stronger magnetic field regarding the Earth magnetic field is mandatory. Nevertheless, exoplanets at the outer region of the habitable zone could be efficiently shielded by an Earth-like magnetic field. The analysis of the star rotation effect on the habitability state due to the SW indicates that exoplanets with an Earth-like magnetic field at $5.0$ au are efficiently protected during extreme space weather conditions if the star rotation period is larger than $2$ days. On the other hand, exoplanets at $2.5$ au requires an intrinsic magnetic field stronger regarding the Earth if the star rotation period is smaller than $10$ days. It should be noted that the rotation period of $\tau$ Boo is $3.3$ days, thus habitability conditions due to the space weather require an exoplanet magnetic field stronger compared to the Earth. That means, habitability conditions may relax for the case of F stars in the spectral range from $F7$ to $F9$ because the rotation period is larger ($10$ days or higher) \citep{Nielsen}. Nevertheless, the habitable zone of $F7$ to $F9$ stars displaces closer to the star, located between $1.1$ to $2.5$ au. Consequently, exoplanets located in the outer region of the habitable zone of $F7$ to $F9$ stars require, at least, a magnetic field similar to the Earth to avoid the direct SW precipitation during CMEs, although it must be stronger if the orbit is closer to the star or the star rotation period is shorter than $10$ days.

\section{Radio emission from exoplanets with an Earth-like magnetosphere}

Radio emission from exoplanet magnetospheres and space weather conditions are closely connected. Radio emission measurements may provide information of the exoplanet magnetic field and, once the characteristics of the exoplanet magnetic field are inferred, insights about the space weather conditions generated by the host star on the exoplanet orbit. This section is dedicated to the analysis of the influence of the space weather conditions, from regular to CME-like, on the radio emission generation, providing simplified new tools for the interpretation of radio telescopes observational data.

The interaction of the SW with a planetary magnetosphere can be analyzed using the analogous of a flow facing a magnetized object, leading to the partial transfer of the flow energy. The transferred energy is transformed to radiation and the radiation power ($P_{disp}$) is proportional to the intercepted flux of the magnetic energy. Thus, following the radio-magnetic Bode’s law, the incident magnetized flow power and the obstacle magnetic field intensity can be used to approximate the radio emission as $P_{w} = \beta [P_{disp}]^{n}$, with $P_{w}$ the radio emission power, $\beta$ the efficiency of dissipated power to radio emission conversion with $n \approx 1$ \citep{Zarka3,Zarka9} and $\beta \approx 2 \cdot 10^{-3} - 10^{-2}$ \citep{Zarka8}.

The power dissipated in the interaction between the SW with the magnetosphere is calculated at the exoplanet day side. Irreversible processes in the interaction convert internal, bulk flow kinetic and magnetic energy into the kinetic energy required to accelerate the electrons along the magnetic field lines, and leading to cyclotron-maser radiation emission by these accelerated electrons. The energy transfer can be evaluated analyzing the energy fluxes of the system. There is a detailed discussion of the flux balance in \citet{Varela6}. The radio emission is calculated using the net magnetic power deposited on the exoplanet day side \citep{Zarka3,Zarka8,Zarka9}:
$$ P_{w} = 2 \cdot 10^{-3} P_{B} = 2 \cdot 10^{-3} \int_{V} \vec{\nabla} \cdot \frac{(\vec{\mathrm{v}}\wedge\vec{B})\wedge\vec{B}}{\mu_{0}} dV $$
with $P_{B}$ the divergence of the magnetic Poynting flux associated with the hot spots of energy transfer in the exoplanet day side and $V$ the volume enclosed between the bow shock nose and the magnetopause.

In the following, the radio emission is calculated during regular and CME-like space weather conditions, modifying the SW dynamic pressure as well as IMF intensity and orientation of the model. First, the effect of the SW dynamic pressure and IMF intensity on the radio emission is analyzed separately. Next, the trends of the radio emission with respect to the SW dynamic pressure and IMF intensity are evaluated together.

\subsection{Effect of the SW dynamic pressure}

This section is dedicated to the study of the exoplanet radio emission generation with respect to the SW density and velocity, hence the SW dynamic pressure. Particular emphasis is dedicated to clarify the link between bow shock compression and radio emission generation.

Figure \ref{5} shows the logarithm of the radio emission power at the exoplanet day side for a set of SW dynamic pressure values increasing the SW velocity (fixed the SW density to $12$ cm$^{-3}$, panel a) and increasing the SW density (fixed the SW velocity to $350$ km/s, panel b) for a star-exoplanet IMF orientation with $|B|_{IMF}=10$ nT. Simulations with $P_{d} < 10$ nPa are analyzed separately due to the effect of the magnetosphere thermal pressure on the magnetopause standoff distance, negligible in the simulations with $P_{d} \geq 10$ nPa \citep{Varela7}.

\begin{figure}[h]
\centering
\resizebox{\hsize}{!}{\includegraphics[width=\columnwidth]{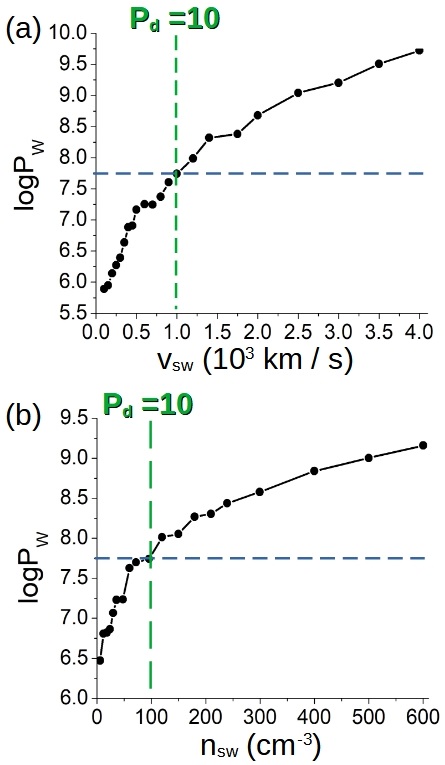}}
\caption{Radio emission power generated in the day side of the exoplanet magnetosphere for a star-exoplanet IMF orientation with $|B|_{IMF}=10$ nT if (a) the SW density is fixed to $12$ cm$^{-3}$ and the SW velocity changes and (b) the SW velocity fixed to $350$ km/s and the SW density changes. The blue dashed horizontal line indicate the radio emission derived from the scaling law by \citet{Zarka8}. The green dashed vertical line indicates the simulations with $P_{d} = 10$ nPa.}
\label{5}
\end{figure}

The radio emission increases from $10^{6}$ to $10^{10}$ W as the SW increases from regular to super CME-like space weather conditions. The order of magnitude of the radio emission power calculated in the simulations is consistent with \citet{Zarka8} scaling (around $6 \cdot 10^{7}$ W) for SW velocity values between $500$ – $1200$ km/s ($P_{d} = 2.5$ – $14$ nPa) and SW density values between $30$ – $120$ cm$^{-3}$ ($P_{d} = 3.1$ – $13.3$ nPa), that is to say, the radio emission values obtained from the simulations and the scaling are similar for regular space weather conditions. If $P_{d} < 2.5$ nPa, the radio emission power is below $10^{7}$ W. For common CME-like conditions ($15 < P_{d} < 40$ nPa) the radio emission power increases up to $6 \cdot 10^{8}$ W. During strong CME-like space weather conditions ($40 < P_{d} < 100$ nPa) the radio emission power reaches $10^{9}$ W. For super CME-like space weather conditions ($P_{d} > 100$ nPa) the radio emission power is $2 \cdot 10^{9}$ W. The enhancement of the radio emission as $P_{d}$ increases is caused by a higher net magnetic power dissipation at the exoplanet day side as the magnetosphere compression intensifies.

Next, the trends of the radio emission with respect to the SW density and velocity are analyzed. Figure \ref{6}, panels a and c, show the fit of the radio emission power to the square value of the SW velocity $P_{w} \propto \Gamma (v_{sw}^{2})^{\alpha}$ if $P_{d} \leq 10$ nPa and $> 10$ nPa, respectively. Figure \ref{6}, panels b and d, show the fit of the radio emission power to the SW density $P_{w} \propto \Gamma (n_{sw})^{\alpha}$ if $P_{d} \leq 10$ nPa and $> 10$ nPa, respectively. The radio emission trends are analyzed separately in the simulations with $P_{d} \leq 10$ nPa and $> 10$ nPa to isolate the effect of the thermal pressure caused by the magnetosphere (for more information please see \citet{Varela7}). The parameters of the data regression are indicated in table \ref{5}.

\begin{figure}[h]
\centering
\resizebox{\hsize}{!}{\includegraphics[width=\columnwidth]{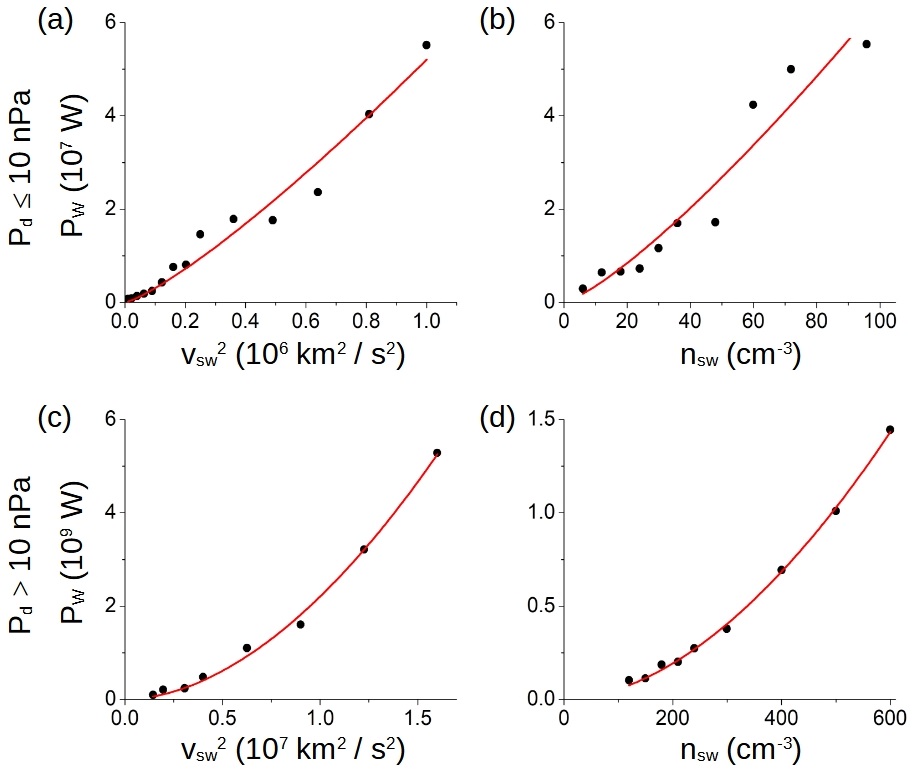}}
\caption{Data regression of the radio emission with respect to the square value of the SW velocity for (a) $P_{d} \leq 10$ and (c) $P_{d} > 10$. Data regression of the radio emission with respect to the SW density for (b) $P_{d} \leq 10$ and (d) $P_{d} > 10$.}
\label{6}
\end{figure}

\begin{table}
\centering
\begin{tabular}{c | c c}
 & $P_{d} \leq 10$ (nPa) & \\ \hline
Regression & $\Gamma$ & $\alpha$ \\ \hline
Velocity & $(2 \pm 3) \cdot 10^5$ & $1.2 \pm 0.1$ \\
Density & $(2 \pm 1) \cdot 10^5$ & $1.3 \pm 0.2$ \\ \hline
 & $P_{d} > 10$ (nPa) & \\ \hline
Velocity & $(3 \pm 4) \cdot 10^{-4}$ & $1.84 \pm 0.08$ \\
Density & $(1.2 \pm 0.3) \cdot 10^4$ & $1.82 \pm 0.04$ \\
\end{tabular}
\caption{Regression parameters in simulations with different SW velocity and density values. (a) Variable SW parameter in the data regression, (b) $\Gamma$ factor and (c) $\alpha$ exponent. Trends in the simulations with $P_{d} \leq 10$ nPa and $P_{d} > 10$ nPa are analyzed separately.}
\label{5}
\end{table}

The data fit finds similar exponents for the regression $P_{w} \propto (v_{sw}^{2})^{\alpha}$ and $P_{w} \propto (n_{sw})^{\alpha}$ if $P_{d} \leq 10$ nPa, that is to say, proportional to the SW dynamic pressure. The scaling of the radio emission with respect to the SW dynamic pressure is stronger in simulations with $P_{d} > 10$ nPa, thus the radio emission generation is further promoted in a compressed magnetosphere. This is explained by the enhancement of the Poynting flux divergence as the magnetopause is located closer to the exoplanet surface. The regression parameters can be compared with the theoretical expression of the radio emission induced by a magnetized flow dominated by the dynamic pressure facing a magnetized obstacle \cite{Zarka8,Zarka9}:
$$P_{W} = \beta \frac{|B_{IMF,\perp}|^{2} B_{ex}^{2/3}}{\mu_{0}^{4/3}} \left( \frac{v_{sw}}{m_{p} n_{sw}} \right)^{1/3} R_{ex}^{2} \pi \frac{2.835}{K^{1/3}}$$
with $B_{IMF,\perp}$ the perpendicular component of the IMF with respect to the flow velocity, $B_{ex}$ the intensity of the magnetic field in the equator of the magnetized obstacle, $\mu_{0}$ the vacuum magnetic permeability and $K =1$-$2$. Here, the intercepted flux of magnetic energy is estimated as $P_{disp} = \epsilon \left( v_{sw} |B_{IMF,\perp}|^{2} / \mu_{0} \right) \pi R_{obs}^{2}$ with $\epsilon = M_{A} / (1+M_{A}^{2})^{1/2}$ ($M_{A}$ Alfvenic Mach number), $R_{obs} = 1.5 R_{mp}$ and $R_{mp} = R_{ex} \left( 2 B_{ex} / (\mu_{0} K n_{sw} v_{sw}^{2}) \right)^{1/6}$. Thus, the theoretical dependency of the radio emission power with the SW velocity is $v_{sw}^{0.33}$ and with the SW density is $n_{sw}^{-0.33}$. The radio emission calculated in the simulations (all dominated by the SW dynamic pressure because $P_{IMF} = 0.09$ nPa) shows a stronger dependency with the SW velocity compared to the theoretical model. Regarding the SW density, the simulations show a direct proportionality with the radio emission, not an inverse proportionality as the theoretical expression predicts. This discrepancy can be explained by the enhancement of the magnetosphere compression and bow shock distortion as the SW dynamic pressure increases, that is to say, the theoretical expression cannot reproduce the effect of the bow shock compression associated with a modification of the energy fluxes, net magnetic power dissipated and divergence of the magnetic Poynting flux in the magnetosphere day side.  Thus, the theoretical scaling law could underestimate the radio emission power generated in exoplanets for space weather conditions leading to a strongly compressed bow shock.

The effect of the SW dynamic pressure on the radio emission generation is highlighted in figure \ref{7}, comparing the divergence of the Poynting flux in the bow shock and magnetopause region for simulations with $v_{sw}=300$ km/s ($P_{d} = 0.9$ nPa) and $v_{sw}=3000$ km/s ($P_{d} = 90$ nPa). The Poynting flux divergence is more than one order of magnitude higher in the simulation with $P_{d} = 90$ nPa, explaining the radio emission enhancement as the SW dynamic pressure increases. It should be noted that the maxima of the Poynting flux divergence is located closer to the exoplanet surface as $P_{d}$ increases because the magnetosphere standoff distance is smaller. In addition, the local maxima of the Poynting flux divergence is displaced towards the South of the magnetosphere in both simulations, determined by the IMF orientation and in particular by the location of the reconnection region. From the observational point of view, radio telescopes may measure a signal with a more localized radio emission maxima as the bow shock compression enhances, although the radio emission maxima should be more diffused as the bow shock compression is weakened.

\begin{figure}[h]
\centering
\resizebox{\hsize}{!}{\includegraphics[width=\columnwidth]{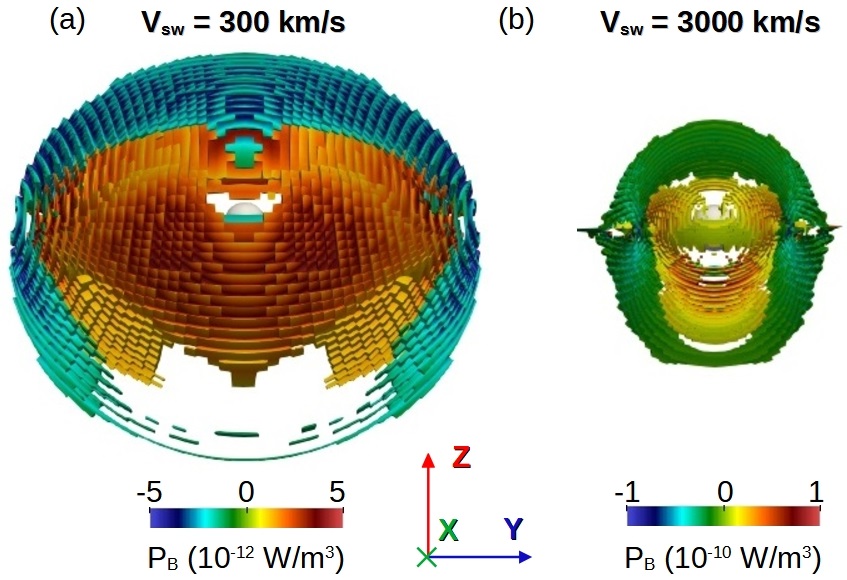}}
\caption{Iso-volume of the Poynting flux divergence in the bow shock and magnetopause region for simulations with (a) $v_{sw}=300$ km/s and (b) $v_{sw}=3000$ km/s. Star-exoplanet IMF orientation with $|B|_{IMF} = 10$ nT and SW density of $12$ cm$^{-3}$. Both panels show plots with the same dimensional scale.} 
\label{7}
\end{figure}

\subsection{Effect of the IMF intensity and orientation}

In this subsection we analyze the effect of the IMF intensity and orientation on the exoplanet radio emission generation. In particular, the role of the reconnection between the IMF and the exoplanet magnetic field is explored, as well as the bow shock formation or dispersion as the SW dynamic pressure or the IMF magnetic pressure dominate, respectively.

\begin{figure}[h]
\centering
\resizebox{\hsize}{!}{\includegraphics[width=\columnwidth]{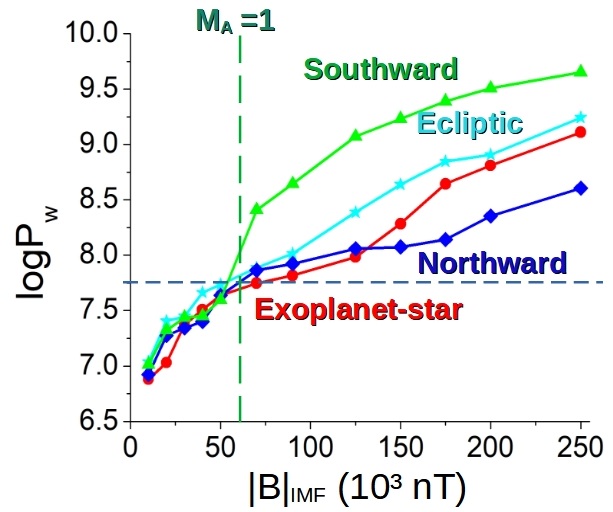}}
\caption{Logarithm of the radio emission power for simulations with $P_{d} = 1.2$ nPa and $|B|_{IMF} = 10 - 250$ nT. IMF orientations: Exoplanet-star (red dots), northward (blue diamonds), southward (green triangle) and ecliptic (cyan stars). The blue dashed horizontal line indicate the radio emission range derived from the scaling law by \citep{Zarka8}. The dark green dashed vertical line indicates the simulations with $M_{A} < 1$ (right) and $M_{A} > 1$ (left).}
\label{8}
\end{figure}

The IMF can induce large distortions in the exoplanet magnetic field, modifying locally the topology of the magnetosphere, particularly in the reconnection regions between the exoplanet magnetic field and the IMF. Figure \ref{8} shows the logarithm of the radio emission fixed $P_{d} = 1.2$ nPa for different IMF orientations (exoplanet-star, northward, southward and ecliptic) and IMF intensities between $10$ and $250$ nT.

The same order of magnitude is obtained for the radio emission power comparing simulation results and \citet{Zarka8} scaling if the IMF intensity is between $20$ - $125$ nT for an exoplanet-star IMF, $10$ - $125$ nT for a northward IMF, $10$ – $50$ nT for a southward IMF and $10$ – $70$ nT for an ecliptic IMF. Consequently, the radio emission calculated in the simulations and the values predicted by the scaling are similar from regular to strong CME-like space weather conditions regarding the IMF intensity. The simulations also predict a radio emission power above $10^{8}$ W during Super CME. The IMF orientation leading to the largest radio emission is the southward IMF, followed by the ecliptic and exoplanet-star IMF. The lowest radio emission is observed for the northward IMF. The variation of the radio emission values regarding the IMF orientation is explained by the location and intensity of the reconnection regions. The southward IMF orientation induces the strongest reconnection, located in the equatorial region of the magnetosphere leading to the smallest magnetopause standoff distance and the largest radio emission. Likewise, the northward IMF orientation causes the lowest radio emission because the reconnection region is located nearby the exoplanet poles and the magnetopause standoff distance is larger regarding the other IMF orientations. It should be noted that the location of the radio emission maxima and the reconnetion regions are concomitant in the simulation, thus the radio emission maxima displaces with the reconnection region as the IMF intensity increases; towards the equatorial region for a southward IMF, the poles for a northward IMF, to the South of the magnetosphere for a star-exoplanet IMF, to the North for a exoplanet-star and tilted to a higher longitude for a IMF oriented in the equatorial plane. 

\begin{figure}[h]
\centering
\resizebox{\hsize}{!}{\includegraphics[width=\columnwidth]{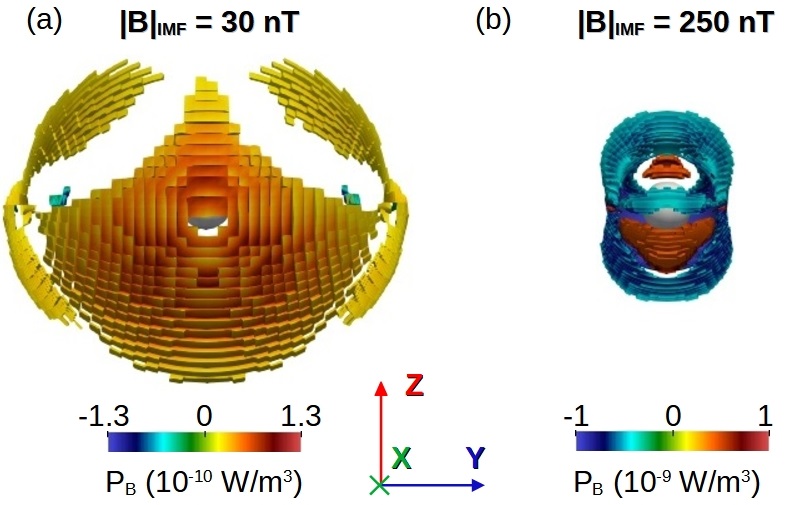}}
\caption{Iso-volume of the Poynting flux divergence in the bow shock and magnetopause region for simulations with (a) $|B|_{IMF}=30$ nT and (b) $|B|_{IMF}=250$ nT. Exoplanet-star IMF orientation and $P_{d} = 1.2$ nPa. Both panels show plots with the same dimensional scale.}
\label{9}
\end{figure}

Figure \ref{9} shows the Poynting flux divergence in the bow shock and magnetopause region for simulations with an exoplanet-star IMF with $|B|_{IMF}=30$ nT (panel a) and $250$ nT (panel b). The radio emission is more than one order of magnitude larger in the simulation with $|B|_{IMF}=250$ nT.

\begin{figure}[h]
\centering
\resizebox{\hsize}{!}{\includegraphics[width=\columnwidth]{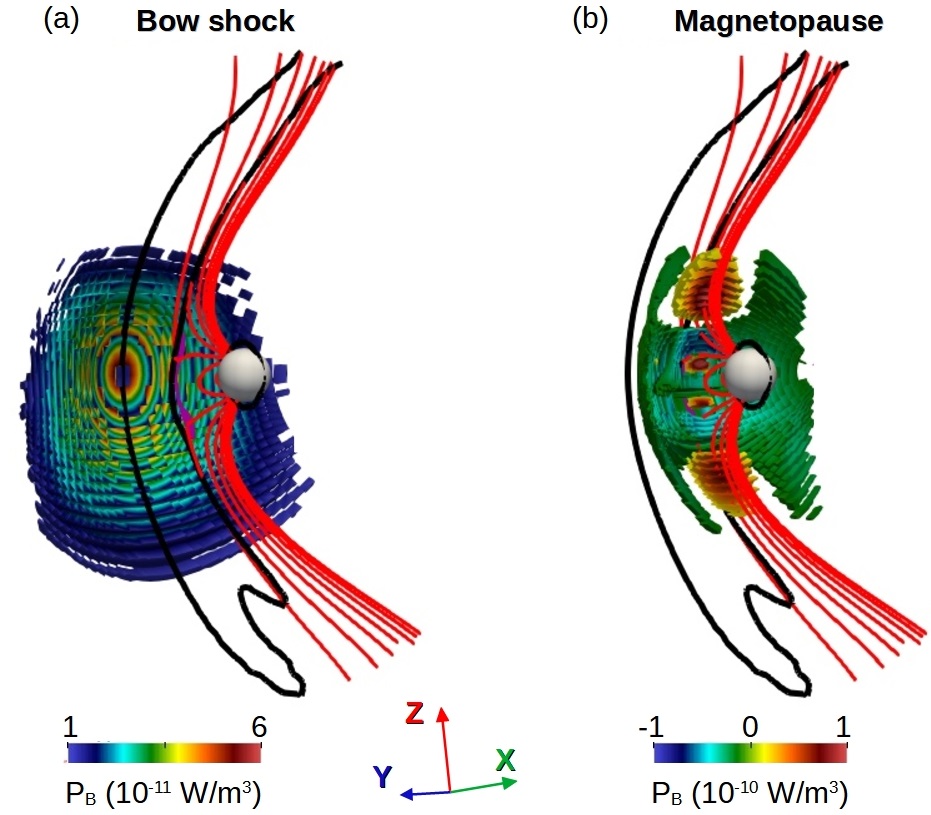}}
\caption{Poynting flux divergence in (a) the bow shock nose and (b) magnetopause reconnection regions. Simulation with southward IMF orientation, $|B|_{IMF} = 30$ nT and $P_{d} = 1.2$ nPa. Black lines indicate the region of the bow shock ($n > 20$ cm$^{-3}$), the red lines the exoplanet magnetic field lines and the pink iso-surface the reconnection region in the XZ plane ($|B| < 5$ nT).} 
\label{10}
\end{figure}

The effect of the IMF orientation on the radio emission is larger in simulations with $|B|_{IMF} \geq 70$ nT. On the other hand, simulations with $|B|_{IMF} < 70$ nT show similar radio emission values for all the IMF orientations. This is explained by the absence of the bow shock in the simulations with $|B|_{IMF} \geq 70$ nT, because the Alfvenic Mach number $M_{A} = v_{sw} / v_{A} < 1$ ($v_{A}$ is the Alfven speed). Simulations with $|B|_{IMF} < 70$ nT ($M_{A} > 1$) lead to the formation of the bow shock, showing two regions with a local maxima of the Poynting flux divergence: 1) the reconnection region between the IMF and the exoplanet magnetic field, 2) the nose of the bow shock where the IMF lines are compressed and bent. Figure \ref{10} shows the radio emission from the bow shock nose, panel a, and the reconnection regions, panel b, for a simulation with southward IMF and $|B|_{IMF} = 30$ nT. The compression and bending of the IMF lines lead to a local maxima of the Poynting flux divergence in the nose of the bow shock. On the other hand, the Poynting flux divergence is larger and more localized in the magnetopause region where the IMF and the exoplanet magnetic field reconnects, closer to the exoplanet surface. Consequently, if the bow shock exists, the Poynting flux divergence in the bow shock depends on the SW dynamic pressure as well, thus the role of the IMF orientation in the radio emission generation is smaller. Radio telescopes may measure a signal with well defined radio emission maxima if the bow shock does not exist, although showing a fast variability of the maxima location as the IMF orientation changes.

\begin{figure}[h]
\centering
\resizebox{\hsize}{!}{\includegraphics[width=\columnwidth]{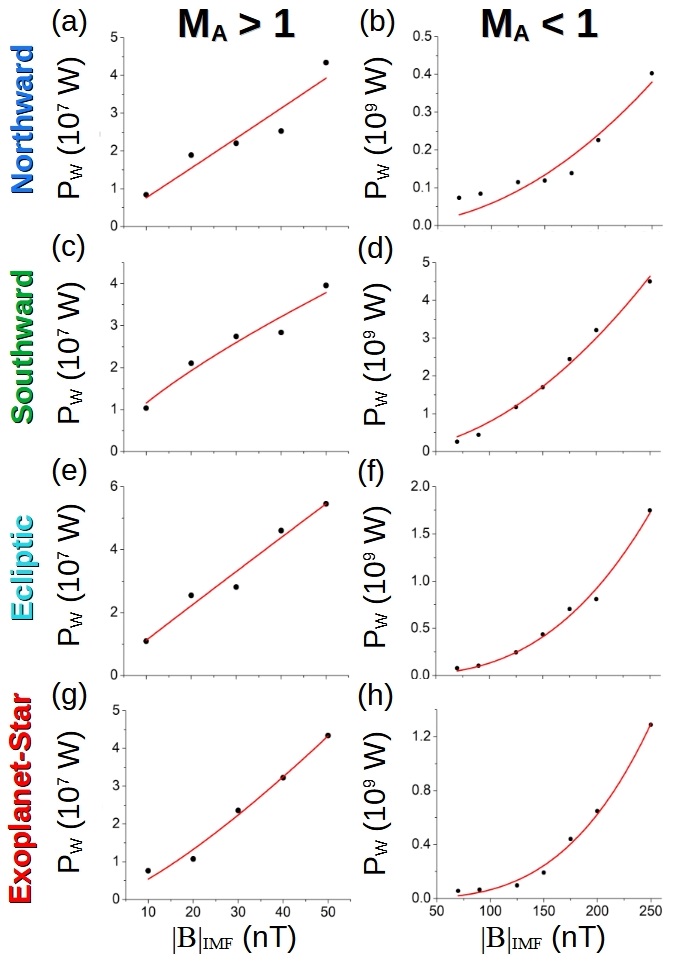}}
\caption{Data fit of the regression $P_{w} \approx \Gamma |B|_{sw}^{\alpha}$ if $|B|_{sw} < 70$ for (a) northward, (c) southward, (e) ecliptic and (g) exoplanet-star IMF. Same data regression if $|B|_{sw} \geq 70$ for (b) northward, (d) southward, (f) ecliptic and (h) exoplanet-star IMF.} 
\label{11}
\end{figure}

Figure \ref{11} and table \ref{6} show the fit of the radio emission values calculated in the simulations using the regression $P_{w} \propto \Gamma |B|_{IMF}^{\alpha}$. It should be noted that the IMF pressure in the simulations with $|B| > 50$ nT is larger than the SW pressure ($P_{IMF} > 1.2$ nPa). In such configurations the theoretical expression of the radio emission is \citep{Zarka8,Zarka9}:
$$P_{W} = \beta \frac{v_{sw} |B_{IMF,\perp}|^{4/3}}{\mu_{0}} R_{ex}^{2} B_{ex}^{2/3} 3.6\pi$$
Here, $R_{mp} = R_{ex} \left( 2 B_{ex} / |B_{IMF,\perp}| \right)^{1/3}$. Thus, the theoretical dependency of the radio emission power with the SW velocity is linear with the $v_{sw}$ and a super linear with the intensity of an IMF perpendicular to the plasma flow. Consequently, the scaling for the simulations with dominant dynamic pressure or dominant IMF pressure must be analyzed separately.

The regression exponents indicate the radio emission dependency with the IMF intensity is weaker in simulations with dominant SW pressure compared to simulations with dominant IMF pressure. This is the opposite tendency with respect to the radio-magnetic scaling law that predicts a stronger $|B|_{IMF}$ trend if the SW pressure is dominant ($|B_{IMF,\perp}|^{2}$). This inconsistency can be explained by the effect of the bow shock compression in the simulations. On the other hand, the regression exponents obtained in simulations with dominant IMF pressure and Southward / Northward IMF orientations are similar to the radio-magnetic scaling law if the dynamic pressure is dominant ($\alpha \approx 2$). That is to say, radio-magnetic scaling law and simulation lead to similar trends if the bow shock does not exist and the IMF is perpendicular to the SW velocity. Consequently, deviations appear if the IMF is unaligned with the exoplanet magnetic field axis and the role of bow shock compression is added in the analysis, effects not included in the radio-magnetic scaling law. In summary, the theoretical scaling law could underestimate the radio emission power generated in exoplanets during space weather conditions leading to the bow shock dispersion.

\begin{table}
\centering
\begin{tabular}{c | c c}
 & $M_{A} > 1$ & \\ \hline
IMF & $\Gamma$ & $\alpha$ \\ \hline
Southward & $(7 \pm 6) \cdot 10^5$ & $1.0 \pm 0.3$ \\
Northward & $(2.1 \pm 0.9) \cdot 10^6$ & $0.74 \pm 0.12$ \\
Exo-star & $(1.6 \pm 0.6) \cdot 10^6$ & $0.98 \pm 0.14$ \\
Ecliptic & $(3 \pm 1) \cdot 10^5$ & $1.29 \pm 0.12$ \\ \hline
 & $M_{A} < 1$ & \\ \hline
Southward & $(5 \pm 9) \cdot 10^3$ & $2.0 \pm 0.3$ \\
Northward & $(1.0 \pm 0.6) \cdot 10^5$ & $1.94 \pm 0.11$ \\
Exo-star & $(3 \pm 3) \cdot 10^2$ & $2.8 \pm 0.12$ \\
Ecliptic & $(2 \pm 2) \cdot 10$ & $3.3 \pm 0.2$
\end{tabular}
\caption{Regression parameters in simulations with different IMF orientations and intensities. IMF orientation (first column), $\Gamma$ factor (second column) and $\alpha$ exponent (third column). The trends in simulations with $M_{A} > 1$ and $M_{A} < 1$ are analyzed separately.}
\label{6}
\end{table}

\subsection{Combined effect of the SW dynamic pressure, IMF intensity and IMF orientation}

The analysis of the combined effect of SW dynamic pressure, IMF intensity and orientation provides an improved approach of the radio emission generation trends, particularly during extreme space weather conditions that melds a large compression of the bow shock and an intense magnetic reconnection.

\begin{figure}[h]
\centering
\resizebox{\hsize}{!}{\includegraphics[width=\columnwidth]{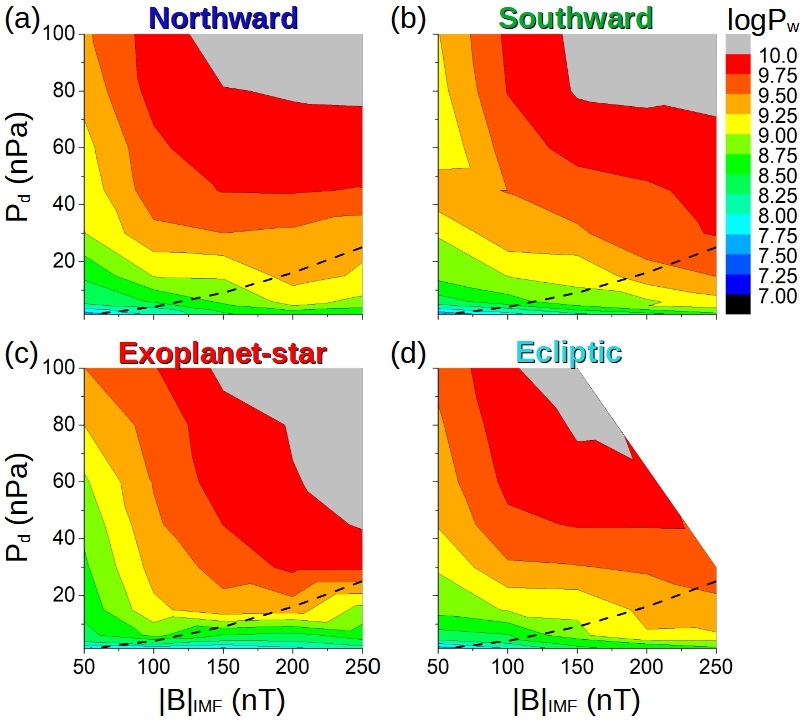}}
\caption{Logarithm of the radio emission with respect to the SW dynamic pressure and IMF intensity for (a) northward, (b) southward, (c) exoplanet-star and (d) ecliptic orientation. The dashed black line indicates the simulations with dominant SW pressure (above the line) and dominant IMF pressure (below the line).} 
\label{12}
\end{figure}

Figure \ref{12} shows the logarithm of the radio emission with respect to the SW dynamic pressure, IMF intensity and orientation for CME-like space weather conditions ($P_{d} = 1.5$ – $100$ nPa and $|B|_{IMF} = 50$ - $250$ nT). It should be noted that the increment of the SW dynamic pressure is the simulations is done by increasing the velocity of the SW, thus the SW density is fixed in the simulations. The radio emission ranges from $3 \cdot 10^{8}$ W for common CME ($20$ nPa and $50$ nT) to above $10^{10}$ W for super CME-like space weather conditions ($100$ nPa and $250$ nT). A large bow shock compression (large SW dynamic pressure) combined with a strong reconnection between IMF and exoplanet magnetic field (IMF intensity is high) lead to a further enhancement of the radio emission. The simulations with large SW dynamic pressure show similar radio emission values independently of the IMF intensity and orientation. On the other hand, the radio emission show larger changes between simulations with different IMF intensity and orientation if the SW dynamic pressure is low. Again, this result is consistent with previous analysis because simulations with low SW dynamic pressure and large IMF (particularly if $M_{A} < 1$) show a larger effect of the IMF intensity and orientation on the radio emission. 

\begin{figure}[h]
\centering
\resizebox{\hsize}{!}{\includegraphics[width=\columnwidth]{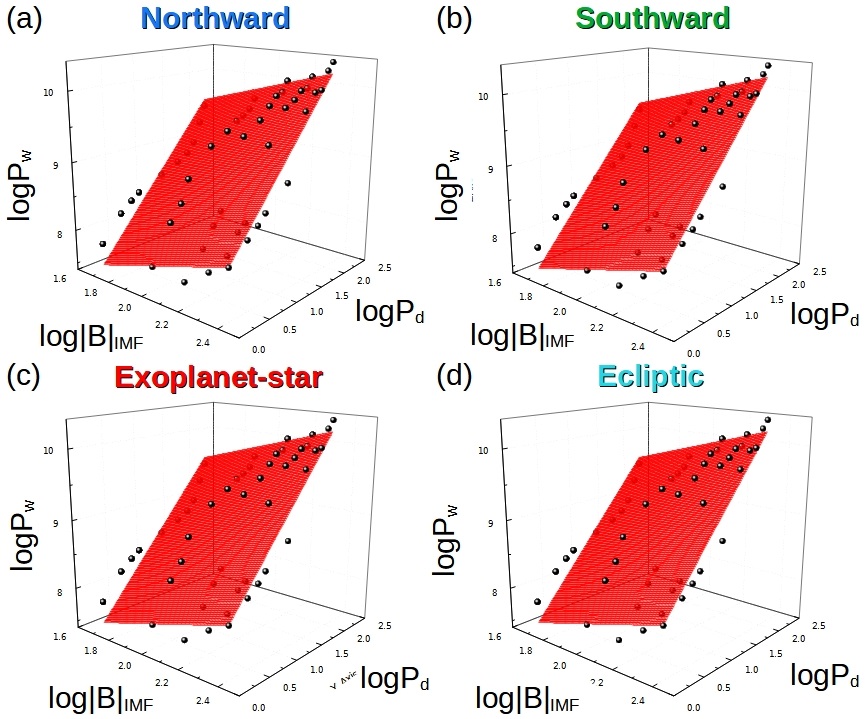}}
\caption{Data fit of the regression $logP_{W} \propto logZ + Mlog(|B|_{IMF}) + Nlog(P_{d})$ for (a) northward, (b)southward, (c) exoplanet-star and (d) ecliptic IMF.} 
\label{13}
\end{figure}

Figure \ref{13} and table \ref{7} indicate the data fit and the parameters of the regression $logP_{W} \propto logZ + Mlog(|B|_{IMF}) + Nlog(P_{d})$, respectively. This expression is derived from $P_{W} \propto Z|B|_{IMF}^{M}P_{d}^{N}$. The data regression includes simulations with dominant SW and dominant IMF pressure because the main part of the space weather conditions analyzed have a dominant SW pressure, indicated by the black dashed line in figure \ref{12} (SW dominant cases above the line).

\begin{table}
\centering
\begin{tabular}{c | c c c}
IMF & $Z$ & $M$ & $N$ \\ \hline
Southward & $5.45 \pm 0.15$ & $1.22 \pm 0.07$ & $0.95 \pm 0.03$ \\
Northward & $5.68 \pm 0.17$ & $1.09 \pm 0.08$ & $0.97 \pm 0.03$ \\
Exoplanet-star & $5.8 \pm 0.3$ & $0.90 \pm 0.12$ & $1.15 \pm 0.05$ \\
Ecliptic & $5.7 \pm 0.2$ & $1.13 \pm 0.07$ & $0.99 \pm 0.03$ \\ \hline
\end{tabular}
\caption{Regression parameters in simulations with different SW dynamic pressure, IMF orientation and intensity. IMF orientation (first column), $Z$ parameter (second column), $M$ parameter (third column) and $N$ parameter (fourth column).}
\label{7}
\end{table}

The regression parameters with respect to the IMF intensity show similar trends compared to simulations with fixed SW dynamic pressure if the bow shock exist ($M \approx 1$ and $\alpha \approx 1$, see table \ref{6} and \ref{7}). On the other hand, the scaling with respect to the SW dynamic pressure is weaker compared to simulations with fixed IMF intensity and orientation ($N \approx 1$ although $\alpha \approx 1.8$ if $P_{d} > 10$ nPa, see table \ref{5} and \ref{7}). Consequently, the simulations analysis indicate the effect of the IMF intensity on the radio emission is similar to the SW dynamic pressure if the bow shock exist and it is strongly compressed. In addition, there is a variation of the radio emission scaling with respect to the IMF orientation up to $20 \%$, pointing out the important role of the IMF orientation on the radio emission generation. If the exponents of the data regression are compared to the radio-magnetic scaling law for a dominant SW dynamic pressure, there is clear deviation showing a weaker trend for $|B|_{IMF}$ ($M \approx 1$ versus $2$) although stronger for $P_{d}$ ($N \approx 1$ versus $0.17$). Such difference is smaller if the regression exponents are compared to the radio-magnetic scaling law for a dominant IMF pressure, showing a similar $|B|_{IMF}$ exponent ($M \approx 1$ versus $1.33$) and a $P_{d}$ exponent $2$ times larger ($N \approx 1$ versus $0.5$). Indeed, the best agreement is obtained if the IMF orientation is Southward ($M = 1.22$ and $N = 0.95$). Consequently, as it was previously discussed, the discrepancy with the radio-magnetic scaling law for the configurations with dominant SW pressure could be caused by the effect of the bow shock compression.

\subsection{Analysis result consequences on the interpretation of radio telescope measurements}

The analysis of the radio emission generated in exoplanet magnetospheres for different space weather conditions provides useful information regarding the variability of the radio emission signal measured by radio telescopes. In addition, an order of magnitude approximation of the radio emission generated by exoplanets with an Earth-like magnetosphere is provided for different space weather conditions.

The combined effect of a strongly compressed bow shock and an intense reconnection between the IMF and the exoplanet magnetic field can lead to a large increase of the radio emission generation. For the case of an exoplanet with an Earth-like magnetic field, the radio emission can increase more than four orders of magnitude comparing regular and extreme space weather conditions (super CME-like events for the case of the Earth).

The simulations indicate that the largest radio emission variability should be observed from exoplanets hosted by stars with large magnetic activity and low SW dynamic pressure, leading to space weather conditions that avoid the formation of the bow shock. The radio emission variation for a given SW dynamic pressure could be close to one order of magnitude regarding the IMF orientation. On the other hand, if the exoplanet is hosted by stars with low magnetic activity although large SW dynamic pressure, the variability of the radio emission with the IMF orientation should be small and mainly induced by changes on the SW dynamic pressure. The variation of the radio emission with the IMF in simulations with bow shock is smaller than a factor $1.5$.

The study also shows that, if the host star generates a SW with large dynamic pressure and an intense IMF, the effect of the IMF orientation should also induce an substantial variability on the radio emission signal even if the bow shock exist, close to a factor $2$. Consequently, a large radio emission variability is linked to unfavorable space weather conditions because the host star magnetic activity is large, leading to a strong reconnection between IMF and exoplanet magnetic field, reducing the magnetopause standoff distance. The same way, a strong radio emission signal combined with a small variability indicates a compressed magnetosphere, that is to say, the SW dynamic pressure generated by the host star is large also reducing the magnetosphere standoff distance.

The simulations scaling shows an underestimation of the exoplanet radio emission by the theoretical scaling for space weather conditions leading to a strongly compressed or vanishing bow shock. Consequently, the radio telescope sensibility required to measure the radio emission generated by terrestrial planets inside the habitable zone of M, K, G and F stars could be lower than expected.

The less restrictive conditions to the exoplanet habitability are linked to a radio emission signal with rather low variability. This is the case for simulations with low SW dynamic pressure and IMF intensity, that is to say, space weather conditions leading to magnetopause standoff distances further away from the exoplanet surface.

The inference of the the magnetic field intensity and topology of exoplanets may need long periods of observational data if one wishes to isolate the effect of the space weather conditions on the radio emission signal. The data filtering could be particularly challenging for the case of exoplanets exposed to recurrent extreme space weather conditions or a dominant IMF pressure, leading to a large radio emission variability. On the other hand, the identification of the magnetic field characteristics for exoplanets facing more benign space weather conditions could be less complex, because the variability of the radio emission data should be smaller.

Once the properties of the exoplanet magnetic field are identified, the analysis of the radio emission time series opens the possibility of tracking the space weather conditions on the exoplanet orbit, providing important information about the host star as the magnetic field or SW dynamic pressure. 

\section{Conclusions and discussion}

Present study is dedicated to analyze the interaction between the stellar wind and exoplanets with an Earth-like magnetosphere hosted by M stars and F star type $\tau$ Boo, in particular the habitability restrictions induced by the sterilizing effect of the stellar wind on the exoplanet surface if the magnetosphere shielding is inefficient. The radio emission generated by exoplanets with an Earth-like magnetosphere is also calculated for different space weather conditions. With that aim, a set of MHD simulations were performed reproducing the interaction of the stellar wind with the exoplanet magnetosphere during regular and extreme space weather conditions.

The simulations results indicate that exoplanets with an Earth-like magnetosphere hosted by a M star at $0.2$ au are protected from the stellar wind during regular and CME-like space weather conditions. This protection holds if the rotation period of the star is $3$ days or larger, although fast rotators can constrain the exoplanet habitability due to the generation of intense and recurrent CME-like events \citep{Aarnio}. Likewise, if the exoplanet orbit is at $0.1$ au, the magnetosphere protection only holds for M stars with a rotation period of $12$ days or larger. On the other hand, if the exoplanet orbit is below $0.1$ au, the magnetic field must be stronger regarding the Earth to avoid the direct impact of the stellar wind at low latitudes, particular during CME-like space weather conditions. It should be noted that the discussion about the properties of the terrestial exoplanet magnetic fields, for example the type of internal magnetic dynamo at different orbits, the spinning rotation speed or the synchronicity with the host star are not explored in this study, although these effects must be consider to improve the accuracy of the predictions \citep{Stevenson}.

If the exoplanet is hosted by a F stars like $\tau$ Boo inside the habitable zone, regular space weather conditions do not impose strong constraint on the habitability. On the other hand, if the exoplanet orbit is close to the inner boundary of the habitable zone ($2.5$ au), an efficient shielding during CME-like space weather conditions requires a stronger magnetic field compared to the Earth. The introduction of the effect of the star rotation in the analysis indicates that the direct precipitation of the SW can occur if the star rotation period is below $10$ days for exoplanets at $2.5$ au during extreme space weather conditions, although for exoplanets at $5$ au the star rotation period must be $2$ days or lower.

The radio emission calculated in simulations with a dynamic pressure between $P_{d} = 2.5 - 14$ nPa shows the same order of magnitude regarding the scaling proposed by \citet{Zarka8}, predicting $7.5 \cdot 10^{7}$ W. That is to say, the radio emission obtained in the simulations is consistent with the scaling during regular and weak CME-like space weather conditions. Likewise, simulations with fixed dynamic pressure ($P_{d} = 1.2$ nPa) also show radio emission values comparable with \citet{Zarka8} scaling if the IMF intensity is in the range of values observed during regular to strong CME-like space weather conditions. In addition, the southward IMF orientation leads to the strongest radio emission and the northward IMF to the lowest. The simulations indicate an enhancement of the radio emission as the stellar wind dynamic pressure and IMF intensity increase. Consequently, radio telescopes may receive a stronger signal from exoplanets hosted by stars with large magnetic activity and intense stellar wind (high SW density and velocity), particularly if the exoplanet orbit is close to the star. Nevertheless, such adverse space weather conditions requires an exoplanet with a intense magnetic field that avoids the collapse of the magnetopause on the exoplanet surface. Such ensemble of space weather and exoplanet magnetic field characteristics are found in Hot Jupiters, reason why the first potential detection of radio emission from an exoplanet involved the Hot Jupiter $\tau$ Boo b \citep{Turner3}. Unfortunately, the radio emission detection from exoplanets hosted by stars with more favorable habitability conditions regarding the space weather inside habitable zone, will require a new generation of radio telescopes with improved resolution and sensibility because the radio emission signal should be several orders of magnitude smaller compared to Hot Jupiters.

The simulations indicate a larger variability of the exoplanet radio emission induced by the IMF orientation if the bow shock does not exist, that is to say, the stellar wind dynamic pressure is low enough and the IMF intensity high enough to be in the parametric range of $M_{A} < 1$. On the other hand, the radio emission variability caused by the IMF orientation is smaller if the bow shock exist ($M_{A} > 1$). That happens because, if the bow shock exist, there is a component of the radio emission linked to the compression and bending of the IMF lines in the nose of the bow shock, mainly dependent on the dynamic pressure of the stellar wind. Thus, the radio emission sources are the bow shock compression and the reconnection site between IMF and exoplanet magnetic field. Consequently, the role of the IMF orientation is smaller with respect to the configurations without bow shock. The implication of this result is that exoplanet magnetospheres routinely perturbed by intense IMF avoiding the formation of the bow shock ($M_{A} < 1$) may show a larger radio emission variability with respect to exoplanet magnetospheres with a bow shock. That is to say, if the exoplanet is hosted by a star with strong magnetic activity although relative low stellar wind dynamic pressure, the radio telescopes may measure a large time variability induced by changes in the IMF orientation, particularly if the magnetosphere erosion leads to a magnetopause located close to the exoplanet surface. Hence, if radio telescopes routinely measure relatively strong and very variable signal, the exoplanet habitability conditions may not be optimal from the point of view of the space weather and the exoplanet magnetic field intensity. The same way, if the host star has a relative weak magnetic activity although generates intense stellar winds (large dynamic pressure), the radio emission detected must be relatively large and show a small variability, pointing out a large compression of the exoplanet magnetosphere and low magnetopause standoff distances, thus the exoplanet habitability state regarding the space weather conditions and the intrinsic magnetic field is less favorable. Therefore, the combination of low radio emission and small variability may indicate the space weather conditions and the intrinsic magnetic field of the exoplanet support lower limitations for the exoplanet habitability, efficiently shield by the magnetosphere from the sterilizing effect of the stellar wind.

The analysis of the simulations combining the effect of the SW dynamic pressure with the IMF orientation and intensity shows radio emission values between $3 \cdot 10^{7}$ W for common CME up to $2 \cdot 10^{10}$ W for super CME. The simulations with large SW dynamic pressure and IMF intensity leads to an enhancement of the radio emission because the bow shock is strongly compressed, the reconnection between the IMF and the exoplanet magnetic field is strong and the magnetopause is located close to the exoplanet surface. The statistical analysis shows similar radio emission trends with respect to the SW dynamic pressure and IMF intensity, although the scaling is slightly affected by the IMF orientation. In particular, the southward IMF leads to the largest IMF intensity dependency, $20 \%$ larger with respect to the SW dynamic pressure trend.

Statistical analysis of the radio emission calculated in the simulations leads to data regression exponents that deviate with respect to the radio-magnetic scaling laws \cite{Zarka8,Zarka9}. Nevertheless, the agreement improves comparing the radio-magnetic scaling law of a configuration with dominant IMF pressure and the data regression for a Southward IMF orientation. Consequently, the trends of radio-magnetic scaling law and simulations are similar if the bow shock does not exist and the IMF is perpendicular to the SW velocity. That means the radio-magnetic scaling laws does not fully capture the effect of the bow shock compression and magnetosphere distortion on the radio emission generation due to the combined effect of the SW and IMF. The scaling law obtained from the simulation is, including the range of exponent values calculated for different IMF orientations:
$$ P_{w} \propto |B|_{IMF}^{(0.9 - 1.22)} P^{(0.95 - 1.15)}_{d}$$
that is to say, the radio-magnetic scaling law for space weather conditions with a dominant SW pressure could overestimate the trend of the IMF intensity ($P_{W} \propto |B_{IMF,\perp}|^{2}$) and underestimate the trend of the SW dynamic pressure ($P_{W} \propto P_{d}^{0.17}$). On the other hand, the prediction of the radio-magnetic scaling law for space weather conditions with a dominant IMF pressure is closer to the simulations scaling regarding the IMF intensity ($P_{W} \propto |B_{IMF,\perp}|^{1.3}$) and the SW dynamic pressure $P_{W} \propto P_{d}^{0.5}$). In summary, the theoretical scaling may underestimate the radio emission generation, particularly with respect to the SW dynamic pressure trend.

A further refinement of the simulations scaling requires an improved description of the model's physics, for example introducing the exoplanet rotation and kinetic effects. Nevertheless, the present study provides a first order approximation of the exoplanet standoff distance and magnetospheric radio emission with respect to the space weather conditions generated by host star.

\begin{acknowledgements}
This work was supported by the project 2019-T1/AMB-13648 founded by the Comunidad de Madrid. The research leading to these results has received funding from the grants ERC WholeSun 810218, Exoplanets A and INSU/PNP. We extend our thanks to CNES for Solar Orbiter, PLATO and Space weather science support and to INSU/PNST for their financial support. This work has been supported by Comunidad de Madrid (Spain) - multiannual agreement with UC3M (“Excelencia para el Profesorado Universitario” - EPUC3M14 ) - Fifth regional research plan 2016-2020. P. Zarka acknowledges funding from the ERC No 101020459 - Exoradio). Data available on request from the authors.
\end{acknowledgements}

\bibliographystyle{aa}
\bibliography{References}

\begin{thebibliography}{101}
\expandafter\ifx\csname natexlab\endcsname\relax\def\natexlab#1{#1}\fi

\bibitem[{{Aarnio, A. N.} {et~al.}(2012){Aarnio, A. N.}, {Matt, S. P.}, \&
  {Stassun, K. G.}}]{Aarnio}
{Aarnio, A. N.}, {Matt, S. P.}, \& {Stassun, K. G.} 2012, The Astrophysical
  Journal, 760, 9

\bibitem[{{Airapetian, V. S.} {et~al.}(2020){Airapetian, V. S.}, {Barnes, R.},
  {Cohen, O.}, {Collinson, G. A.}, {Danchi, W. C.}, {Dong, C. F.}, {Del Genio,
  A. D.}, {France, K.}, {Garcia-Sage, K.}, {Glocer, A.}, \&
  et~al.}]{Airapetian}
{Airapetian, V. S.}, {Barnes, R.}, {Cohen, O.}, {et~al.} 2020, International
  Journal of Astrobiology, 19, 136–194

\bibitem[{{Bailey, J. D.}(2014)}]{Bailey}
{Bailey, J. D.} 2014, A\&A, 568, A38

\bibitem[{{Barnes}(2017)}]{Barnes}
{Barnes}, R. 2017, Celestial Mechanics and Dynamical Astronomy, 129, 509

\bibitem[{{Benjamin J. S.} {et~al.}(2020){Benjamin J. S.}, {Pope, B. J. S.},
  {Bedell, M.}, {Callingham, J. R.}, {Vedantham, H. K.}, {Snellen, T.},
  {Price-Whelan, A. M.}, \& {Shimwell, T. W.}}]{Pope}
{Benjamin J. S.}, {Pope, B. J. S.}, {Bedell, M.}, {et~al.} 2020, The
  Astrophysical Journal, 890, L19

\bibitem[{{Brun, A. S.} {et~al.}(2022){Brun, A. S.}, {Strugarek, A.}, {Noraz,
  Q.}, {Perri, B.}, {Varela, J.}, {Augustson, K.}, {Charbonneau, P.}, \&
  {Toomre, J.}}]{Brun}
{Brun, A. S.}, {Strugarek, A.}, {Noraz, Q.}, {et~al.} 2022, The Astrophysical
  Journal, 926, 21

\bibitem[{{Cane, H. V.} \& {Richardson, I. G.}(2003)}]{Cane2}
{Cane, H. V.} \& {Richardson, I. G.} 2003, Journal of Geophysical Research:
  Space Physics, 108, 1

\bibitem[{{Carilli} \& {Rawlings}(2004)}]{Carilli}
{Carilli}, C.~L. \& {Rawlings}, S. 2004, New Astronomy Reviews, 48, 979

\bibitem[{{Cuntz, M.} \& {Guinan, E. F.}(2016)}]{Cuntz}
{Cuntz, M.} \& {Guinan, E. F.} 2016, The Astrophysical Journal, 827, 79

\bibitem[{{Facskó, G.} {et~al.}(2016){Facskó, G.}, {Honkonen, I.},
  {Živković, T.}, {Palin, L.}, {Kallio, E.}, {Ågren, K.}, {Opgenoorth, H.},
  {Tanskanen, E. I.}, \& {Milan, S.}}]{Facsko}
{Facskó, G.}, {Honkonen, I.}, {Živković, T.}, {et~al.} 2016, Space Weather,
  14, 351

\bibitem[{{Fares, R.} {et~al.}(2009){Fares, R.}, {Donati, J.-F.}, {Moutou, C.},
  {Bohlender, D.}, {Catala, C.}, {Deleuil, M.}, {Shkolnik, E.}, {Cameron, A.
  C.}, {Jardine, M. M.}, \& {Walker, G. A. H.}}]{Fares}
{Fares, R.}, {Donati, J.-F.}, {Moutou, C.}, {et~al.} 2009, Monthly Notices of
  the Royal Astronomical Society, 398, 1383

\bibitem[{{Fares, R.} {et~al.}(2013){Fares, R.}, {Moutou, C.}, {Donati, J.},
  {Catala, C.}, {Shkolnik, E. L.}, {Jardine, M. M.}, {Cameron, A. C.}, \&
  {Deleuil, M.}}]{Fares2}
{Fares, R.}, {Moutou, C.}, {Donati, J.}, {et~al.} 2013, Monthly Notices of the
  Royal Astronomical Society, 435, 1451

\bibitem[{{Gallet, F.} {et~al.}(2017){Gallet, F.}, {Charbonnel, C.}, {Amard,
  L.}, {Brun, S.}, {Palacios, A.}, \& {Mathis, S.}}]{Gallet}
{Gallet, F.}, {Charbonnel, C.}, {Amard, L.}, {et~al.} 2017, A\&A, 597, A14

\bibitem[{{Garraffo, C.} {et~al.}(2016){Garraffo, C.}, {Drake, J. J.}, \&
  {Cohen, O.}}]{Garraffo}
{Garraffo, C.}, {Drake, J. J.}, \& {Cohen, O.} 2016, The Astrophysical Journal,
  833, L4

\bibitem[{{Garraffo, C.} {et~al.}(2017){Garraffo, C.}, {Drake, J. J.}, {Cohen,
  O.}, {Alvarado-G{\'{o}}mez, J. D.}, \& {Moschou, S. P.}}]{Garraffo2}
{Garraffo, C.}, {Drake, J. J.}, {Cohen, O.}, {Alvarado-G{\'{o}}mez, J. D.}, \&
  {Moschou, S. P.} 2017, The Astrophysical Journal, 843, L33

\bibitem[{{Gombosi}(1994)}]{Gombosi}
{Gombosi}, T.~I. 1994, Gaskinetic Theory, Cambridge Atmospheric and Space
  Science Series (Cambridge University Press)

\bibitem[{{González Hernández, I.} {et~al.}(2014){González Hernández, I.},
  {Komm, R.}, {Pevtsov, A.}, \& {Leibacher, J.}}]{Gonzalez}
{González Hernández, I.}, {Komm, R.}, {Pevtsov, A.}, \& {Leibacher, J.} 2014,
  Solar Origins of Space Weather and Space Climate (Springer-Verlag New York)

\bibitem[{{Grie\ss{}meier, J.-M.} {et~al.}(2005){Grie\ss{}meier, J.-M.},
  {Stadelmann, A.}, {Motschmann, U.}, {Belisheva, N.K.}, {Lammer, H.}, \&
  {Biernat, H.K.}}]{Griemeier4}
{Grie\ss{}meier, J.-M.}, {Stadelmann, A.}, {Motschmann, U.}, {et~al.} 2005,
  Astrobiology, 5, 587

\bibitem[{{Grie\ss{}meier, J.-M.} {et~al.}(2004){Grie\ss{}meier, J.-M.},
  {Stadelmann, A.}, {Penz, T.}, {Lammer, H.}, {Selsis, F.}, {Ribas, I.},
  {Guinan, E. F.}, {Motschmann, U.}, {Biernat, H. K.}, \& {Weiss, W.
  W.}}]{Griemeier3}
{Grie\ss{}meier, J.-M.}, {Stadelmann, A.}, {Penz, T.}, {et~al.} 2004, A\& A,
  425, 753

\bibitem[{{Hapgood, M.}(2019)}]{Hapgood}
{Hapgood, M.} 2019, Space Weather, 17, 950

\bibitem[{{Hess} \& {Zarka}(2011)}]{Hess}
{Hess}, S. L.~G. \& {Zarka}, P. 2011, Astron. Astrophys., 531, A29

\bibitem[{{Howard, R.A.}(2006)}]{Howard}
{Howard, R.A.} 2006, A Historical Perspective on Coronal Mass Ejections
  (American Geophysical Union (AGU)), 7--13

\bibitem[{{Hu} \& {Yang}(2014)}]{Hu2}
{Hu}, Y. \& {Yang}, J. 2014, Proceedings of the National Academy of Sciences,
  111, 629

\bibitem[{{Jakosky, B. M.} {et~al.}(2015){Jakosky, B. M.}, {Grebowsky, J. M.},
  {Luhmann, J. G.}, \& {Brain, D. A.}}]{Jakosky}
{Jakosky, B. M.}, {Grebowsky, J. M.}, {Luhmann, J. G.}, \& {Brain, D. A.} 2015,
  Geophysical Research Letters, 42, 8791

\bibitem[{{Jia} {et~al.}(2015){Jia}, {Slavin}, {Gombosi}, {Daldorff}, {Toth},
  \& {Holst}}]{2015JGRA..120.4763J}
{Jia}, X., {Slavin}, J.~A., {Gombosi}, T.~I., {et~al.} 2015, Journal of
  Geophysical Research (Space Physics), 120, 4763

\bibitem[{{Johnstone, C. P.} {et~al.}(2015{\natexlab{a}}){Johnstone, C. P.},
  {G\"udel, M.}, {Brott, I.}, \& {L\"uftinger, T.}}]{Johnstone}
{Johnstone, C. P.}, {G\"udel, M.}, {Brott, I.}, \& {L\"uftinger, T.}
  2015{\natexlab{a}}, A\& A, 577, A28

\bibitem[{{Johnstone, C. P.} {et~al.}(2015{\natexlab{b}}){Johnstone, C. P.},
  {G\"udel, M.}, {L\"uftinger, T.}, {Toth, G.}, \& {Brott, I.}}]{Johnstone2}
{Johnstone, C. P.}, {G\"udel, M.}, {L\"uftinger, T.}, {Toth, G.}, \& {Brott,
  I.} 2015{\natexlab{b}}, A\& A, 577, A27

\bibitem[{{Kabin} {et~al.}(2008){Kabin}, {Heimpel}, {Rankin}, {Aurnou},
  {G{\'o}mez-P{\'e}rez}, {Paral}, {Gombosi}, {Zurbuchen}, {Koehn}, \&
  {DeZeeuw}}]{2008Icar..195....1K}
{Kabin}, K., {Heimpel}, M.~H., {Rankin}, R., {et~al.} 2008, Icarus, 195, 1

\bibitem[{{Kaiser} \& {Desch}(1984)}]{Kaiser}
{Kaiser}, M.~L. \& {Desch}, M.~D. 1984, Reviews of Geophysics, 22, 373

\bibitem[{{Kasting, J. F.} {et~al.}(1993){Kasting, J. F.}, {Whitmire, D. P.},
  \& {Reynolds, R. T.}}]{Kasting}
{Kasting, J. F.}, {Whitmire, D. P.}, \& {Reynolds, R. T.} 1993, Icarus, 101,
  108

\bibitem[{{Khodachenko} {et~al.}(2007){Khodachenko}, {Ribas}, {Lammer},
  {Griessmeier}, {Leitner}, {Selsis}, {Eiroa}, {Hanslmeier}, {Biernat},
  {Farrugia}, \& {Rucker}}]{Khodachenko}
{Khodachenko}, M.~L., {Ribas}, I., {Lammer}, H., {et~al.} 2007, Astrobiology,
  7, 167

\bibitem[{{Kilpua, E.K.J.} {et~al.}(2019){Kilpua, E.K.J.}, {Lugaz, N.}, {Mays,
  M. L.}, \& {Temmer, M.}}]{Kilpua}
{Kilpua, E.K.J.}, {Lugaz, N.}, {Mays, M. L.}, \& {Temmer, M.} 2019, Space
  Weather, 17, 498

\bibitem[{{Kopparapu, R. K.} {et~al.}(2013){Kopparapu, R. K.}, {Ramirez, R.},
  {Kasting, J. F.}, {Eymet, V.}, {Robinson, T. D.}, {Mahadevan, S.}, {Terrien,
  R. C.}, {Domagal-Goldman, S.}, {Meadows, V.}, \& {Deshpande, R.}}]{Kopparapu}
{Kopparapu, R. K.}, {Ramirez, R.}, {Kasting, J. F.}, {et~al.} 2013, The
  Astrophysical Journal, 765, 131

\bibitem[{{Kopparapu, R. K.} {et~al.}(2014){Kopparapu, R. K.}, {Ramirez, R.
  M.}, {SchottelKotte, J.}, {Kasting, J. F.}, {Domagal-Goldman, S.}, \& {Eymet,
  V.}}]{Kopparapu2}
{Kopparapu, R. K.}, {Ramirez, R. M.}, {SchottelKotte, J.}, {et~al.} 2014, The
  Astrophysical Journal, 787, L29

\bibitem[{{Lammer} {et~al.}(2007){Lammer}, {Lichtenegger}, {Kulikov},
  {Griessmeier}, {Terada}, {Erkaev}, {Nikolai}, {Biernat}, {Khodachenko},
  {Ribas}, {Penz}, \& {Selsis}}]{Lammer}
{Lammer}, H., {Lichtenegger}, I.~M., {Kulikov}, Y.~N., {et~al.} 2007,
  Astrobiology, 7, 85

\bibitem[{{Lamy} {et~al.}(2017){Lamy}, {Zarka}, {Cecconi}, {Klein}, {Masson},
  {Denis}, {Coffre}, \& {Viou}}]{Lamy}
{Lamy}, L., {Zarka}, P., {Cecconi}, B., {et~al.} 2017, 455

\bibitem[{{Leconte} {et~al.}(2015){Leconte}, {Wu}, {Menou}, \&
  {Murray}}]{Leconte}
{Leconte}, J., {Wu}, H., {Menou}, K., \& {Murray}, N. 2015, Science, 347, 632

\bibitem[{{Leitzinger} {et~al.}(2020){Leitzinger}, {Odert}, {Greimel}, {Vida},
  {Kriskovics}, {Guenther}, {Korhonen}, {Koller}, {Hanslmeier}, {Kovari}, \&
  {Lammer}}]{Leitzinger}
{Leitzinger}, M., {Odert}, P., {Greimel}, R., {et~al.} 2020, Monthly Notices of
  the Royal Astronomical Society, 493, 4570

\bibitem[{{Linsky, J.}(2019)}]{Linsky}
{Linsky, J.} 2019, Space Weather: The Effects of Host Star Flares on
  Exoplanets, Vol. 955 (Springer International Publishing), 229--242

\bibitem[{{Low, B. C.}(2001)}]{Low}
{Low, B. C.} 2001, Journal of Geophysical Research: Space Physics, 106, 25141

\bibitem[{{Lugaz, N.} {et~al.}(2015){Lugaz, N.}, {Farrugia, C. J.}, {Huang,
  C.-L.}, \& {Spence, H. E.}}]{Lugaz}
{Lugaz, N.}, {Farrugia, C. J.}, {Huang, C.-L.}, \& {Spence, H. E.} 2015,
  Geophysical Research Letters, 42, 4694

\bibitem[{{Lundin, R.} {et~al.}(2007){Lundin, R.}, {Lammer, H.}, \& {Ribas,
  I.}}]{Lundin}
{Lundin, R.}, {Lammer, H.}, \& {Ribas, I.} 2007, Space Science Reviews, 129,
  245

\bibitem[{{Marsden, S. C.} {et~al.}(2014){Marsden, S. C.}, {Petit, P.},
  {Jeffers, S. V.}, {Morin, J.}, {Fares, R.}, {Reiners, A.}, {do Nascimento,
  J.-D.}, {Auriere, M.}, {Bouvier, J.}, {Carter, B. D.}, {Catala, C.},
  {Dintrans, B.}, {Donati, J.-F.}, {Gastine, T.}, {Jardine, M.},
  {Konstantinova-Antova, R.}, {Lanoux, J.}, {Lignieres, F.}, {Morgenthaler,
  A.}, {Ramirez-Velez, J. C.}, {Théado, S.}, {Van Grootel, V.}, \& {the BCool
  Collaboration}}]{Marsden}
{Marsden, S. C.}, {Petit, P.}, {Jeffers, S. V.}, {et~al.} 2014, Monthly Notices
  of the Royal Astronomical Society, 444, 3517

\bibitem[{{Mathur, S.} {et~al.}(2014){Mathur, S.}, {Garc\'{\i}a, R. A.},
  {Ballot, J.}, {Ceillier, T.}, {Salabert, D.}, {Metcalfe, T. S.}, {R\'egulo,
  C.}, {Jim\'enez, A.}, \& {Bloemen, S.}}]{Mathur}
{Mathur, S.}, {Garc\'{\i}a, R. A.}, {Ballot, J.}, {et~al.} 2014, A\& A, 562,
  A124

\bibitem[{{Matt, S. P.} {et~al.}(2015){Matt, S. P.}, {Brun, A. S.}, {Baraffe,
  I.}, {Bouvier, T.}, \& {Chabrier, G.}}]{Matt}
{Matt, S. P.}, {Brun, A. S.}, {Baraffe, I.}, {Bouvier, T.}, \& {Chabrier, G.}
  2015, The Astrophysical Journal, 799, L23

\bibitem[{{Menvielle, M.} \& {Berthelier, A.}(1991)}]{Menvielle}
{Menvielle, M.} \& {Berthelier, A.} 1991, Reviews of Geophysics, 29, 415

\bibitem[{{Mignone} {et~al.}(2007){Mignone}, {Bodo}, {Massaglia}, {Matsakos},
  {Tesileanu}, {Zanni}, \& {Ferrari}}]{Mignone}
{Mignone}, A., {Bodo}, G., {Massaglia}, S., {et~al.} 2007, ApJS, 170, 228

\bibitem[{{Moore, T. E.} \& {Khazanov, G. V.}(2010)}]{Moore}
{Moore, T. E.} \& {Khazanov, G. V.} 2010, Journal of Geophysical Research:
  Space Physics, 115, A00J13

\bibitem[{{Nan} {et~al.}(2011){Nan}, {Li}, {Jin}, {Wang}, {Zhu}, {Zhu},
  {Zhang}, {Yue}, \& {Qian}}]{Nan}
{Nan}, R., {Li}, D., {Jin}, C., {et~al.} 2011, International Journal of Modern
  Physics D, 20, 989

\bibitem[{{Neugebauer} \& {Goldstein}(1997)}]{Neugebauer}
{Neugebauer}, M. \& {Goldstein}, R. 1997, Washington DC American Geophysical
  Union Geophysical Monograph Series, 99, 245

\bibitem[{{Newton, E. R.} {et~al.}(2018){Newton, E. R.}, {Mondrik, N.}, {Irwin,
  J.}, {Winters, J. G.}, \& {Charbonneau, D.}}]{Newton}
{Newton, E. R.}, {Mondrik, N.}, {Irwin, J.}, {Winters, J. G.}, \& {Charbonneau,
  D.} 2018, The Astronomical Journal, 156, 217

\bibitem[{{Nielsen, M. B.} {et~al.}(2013){Nielsen, M. B.}, {Gizon, L.},
  {Schunker, H.}, \& {Karoff, C.}}]{Nielsen}
{Nielsen, M. B.}, {Gizon, L.}, {Schunker, H.}, \& {Karoff, C.} 2013, A\&A, 55,
  L10

\bibitem[{{Odstrcil, D.} \& {Pizzo, V. J.}(1999)}]{Odstrcil2}
{Odstrcil, D.} \& {Pizzo, V. J.} 1999, Journal of Geophysical Research: Space
  Physics, 104, 483

\bibitem[{{Odstrcil, D.} {et~al.}(2004){Odstrcil, D.}, {Riley, P.}, \& {Zhao,
  X. P.}}]{Odstrcil}
{Odstrcil, D.}, {Riley, P.}, \& {Zhao, X. P.} 2004, Journal of Geophysical
  Research: Space Physics, 109

\bibitem[{{P\'erez-Torres, M.} {et~al.}(2021){P\'erez-Torres, M.}, {G\'omez, J.
  F.}, {Ortiz, J. L.}, {Leto, P.}, {Anglada, G.}, {G\'omez, J. L.},
  {Rodr\'{\i}guez, E.}, {Trigilio, C.}, {Amado, P. J.}, {Alberdi, A.},
  {Anglada-Escud\'e, G.}, {Osorio, M.}, {Umana, G.}, {Berdi\~nas, Z.},
  {L\'opez-Gonz\'alez, M. J.}, {Morales, N.}, {Rodr\'{\i}guez-L\'opez, C.}, \&
  {Chibueze, J.}}]{Torres}
{P\'erez-Torres, M.}, {G\'omez, J. F.}, {Ortiz, J. L.}, {et~al.} 2021, A\&A,
  645, A77

\bibitem[{{Perger, M.} {et~al.}(2021){Perger, M.}, {Ribas, I.},
  {Anglada-Escud\'e, G.}, {Morales, J. C.}, {Amado, P. J.}, {Caballero, J. A.},
  {Quirrenbach, A.}, {Reiners, A.}, {B\'ejar, V. J. S.}, {Dreizler, S.},
  {Galad\'{\i}-Enr\'{\i}quez, D.}, {Hatzes, A. P.}, {Henning, Th.}, {Jeffers,
  S. V.}, {Kaminski, A.}, {K\"urster, M.}, {Lafarga, M.}, {Montes, D.},
  {Pall\'e, E.}, {Rodr\'{\i}guez-L\'opez, C.}, {Schweitzer, A.}, {Zapatero
  Osorio, M. R.}, \& {Zechmeister, M.}}]{Perger}
{Perger, M.}, {Ribas, I.}, {Anglada-Escud\'e, G.}, {et~al.} 2021, A\&A, 649,
  L12

\bibitem[{{Popinchalk, M.} {et~al.}(2021){Popinchalk, M.}, {Faherty, J. K.},
  {Kiman, R.}, {Gagne, J.}, {Curtis, J. L.}, {Angus, R.}, {Cruz, K. L.}, \&
  {Rice, E. L.}}]{Popinchalk}
{Popinchalk, M.}, {Faherty, J. K.}, {Kiman, R.}, {et~al.} 2021, The
  Astrophysical Journal, 916, 77

\bibitem[{{Poppe, B.B.} \& {Jorden, K.P.}(2006)}]{Poppe}
{Poppe, B.B.} \& {Jorden, K.P.} 2006, Sentinels of the Sun: Forecasting Space
  Weather (Johnson Books)

\bibitem[{{Raeder, J.} {et~al.}(2001){Raeder, J.}, {McPherron, R. L.}, {Frank,
  L. A.}, {Kokubun, S.}, {Lu, G.}, {Mukai, T.}, {Paterson, W. R.}, {Sigwarth,
  J. B.}, {Singer, H. J.}, \& {Slavin, J. A.}}]{Raeder2}
{Raeder, J.}, {McPherron, R. L.}, {Frank, L. A.}, {et~al.} 2001, Journal of
  Geophysical Research: Space Physics, 106, 381

\bibitem[{{Regnault, F.} {et~al.}(2020){Regnault, F.}, {Janvier, M.},
  {Demoulin, P.}, {Auchere, F.}, {Strugarek, A.}, {Dasso, S.}, \& {Nous,
  C.}}]{Regnault}
{Regnault, F.}, {Janvier, M.}, {Demoulin, P.}, {et~al.} 2020, Journal of
  Geophysical Research: Space Physics, 125, e2020JA028150

\bibitem[{{Ricci} {et~al.}(2018){Ricci}, {Liu}, {Isella}, \& {Li}}]{Ricci2}
{Ricci}, L., {Liu}, S.-F., {Isella}, A., \& {Li}, H. 2018, Astrophys. J., 853,
  110

\bibitem[{{Saffe, C.} {et~al.}(2005){Saffe, C.}, {G\'omez, M.}, \& {Chavero,
  C.}}]{Saffe}
{Saffe, C.}, {G\'omez, M.}, \& {Chavero, C.} 2005, A\& A, 443, 609

\bibitem[{{Salman, T. M.} {et~al.}(2018){Salman, T. M.}, {Lugaz, N.},
  {Farrugia, C. J.}, {Winslow, R. M.}, {Galvin, A. B.}, \& {Schwadron, N.
  A.}}]{Salman}
{Salman, T. M.}, {Lugaz, N.}, {Farrugia, C. J.}, {et~al.} 2018, Space Weather,
  16, 2004

\bibitem[{{Sato, S.} {et~al.}(2014){Sato, S.}, {Cuntz, M.}, {Guerra Olvera, C.
  M.}, {Jack, D.}, \& {Schroder, K.-P.}}]{Sato}
{Sato, S.}, {Cuntz, M.}, {Guerra Olvera, C. M.}, {Jack, D.}, \& {Schroder,
  K.-P.} 2014, International Journal of Astrobiology, 13, 244–258

\bibitem[{{Scalo, J.} {et~al.}(2007){Scalo, J.}, {Kaltenegger, L.}, {Segura,
  A.}, {Fridlund, M.}, {Ribas, I.}, {Kulikov, Y. N.}, {Grenfell, J. L.},
  {Rauer, H.}, {Odert, P.}, {Leitzinger, M.}, {Selsis, F.}, {Khodachenko, M.
  L.}, {Eiroa, C.}, {Kasting, J.}, \& {Lammer, H.}}]{Scalo}
{Scalo, J.}, {Kaltenegger, L.}, {Segura, A.}, {et~al.} 2007, Astrobiology, 7,
  85

\bibitem[{{Seach, J. M.} {et~al.}(2020){Seach, J. M.}, {Marsden, S. C.},
  {Carter, D. B.}, {Neiner, C.}, {Folsom, C. P.}, {Mengel, M. W.}, {Oksala, M.
  E.}, \& {Buysschaert, B.}}]{Seach}
{Seach, J. M.}, {Marsden, S. C.}, {Carter, D. B.}, {et~al.} 2020, Monthly
  Notices of the Royal Astronomical Society, 494, 5682

\bibitem[{{See, V.} {et~al.}(2019){See, V.}, {Matt, S. P.}, {Folsom, C. P.},
  {Saikia, S. B.}, {Donati, J.-F.}, {Fares, R.}, {Finley, A. J.}, {Hebrard, E.
  M.}, {Jardine, M. M.}, {Jeffers, S. V.}, {Lehmann, L. T.}, {Marsden, S. C.},
  {Mengel, M. W.}, {Morin, J.}, {Petit, P.}, {Vidotto, A. A.}, \& {Waite, I.
  A.}}]{See}
{See, V.}, {Matt, S. P.}, {Folsom, C. P.}, {et~al.} 2019, The Astrophysical
  Journal, 876, 118

\bibitem[{{Shields, Aomawa L.} {et~al.}(2016){Shields, Aomawa L.}, {Ballard,
  S.}, \& {Johnson, J. A.}}]{Shields}
{Shields, Aomawa L.}, {Ballard, S.}, \& {Johnson, J. A.} 2016, Physics Reports,
  663, 1

\bibitem[{{Shoda, M.} {et~al.}(2020){Shoda, M.}, {Suzuki, T. K.}, {Matt, S.
  P.}, {Cranmer, S. R.}, {Vidotto, A. A.}, {Strugarek, A.}, {See, V.},
  {R{\'{e}}ville, V.}, {Finley, A. J.}, \& {Brun, A. S.}}]{Shoda}
{Shoda, M.}, {Suzuki, T. K.}, {Matt, S. P.}, {et~al.} 2020, The Astrophysical
  Journal, 896, 123

\bibitem[{{Shulyak, D.} {et~al.}(2017){Shulyak, D.}, {Reiners, A.}, {Engeln,
  A.}, {Malo, L.}, {Yadav, R.}, {Morin, J.}, \& {Kochukhov, O.}}]{Shulyak}
{Shulyak, D.}, {Reiners, A.}, {Engeln, A.}, {et~al.} 2017, Nature Astronomy, 1,
  0184

\bibitem[{{Shulyak, D.} {et~al.}(2019){Shulyak, D.}, {Reiners, A.}, {Nagel,
  E.}, {Tal-Or, L.}, {Caballero, J. A.}, {Zechmeister, M.}, {B\'ejar, V. J.
  S.}, {Cort\'es-Contreras, M.}, {Martin, E. L.}, {Kaminski, A.}, {Ribas, I.},
  {Quirrenbach, A.}, {Amado, P. J.}, {Anglada-Escud\'e, G.}, {Bauer, F. F.},
  {Dreizler, S.}, {Guenther, E. W.}, {Henning, T.}, {Jeffers, S. V.},
  {K\"urster, M.}, {Lafarga, M.}, {Montes, D.}, {Morales, J. C.}, \& {Pedraz,
  S.}}]{Shulyak2}
{Shulyak, D.}, {Reiners, A.}, {Nagel, E.}, {et~al.} 2019, A\& A, 626, A86

\bibitem[{{Stevenson, D. J.}(2003)}]{Stevenson}
{Stevenson, D. J.} 2003, Earth and Planetary Science Letters, 208, 1

\bibitem[{{Strugarek} {et~al.}(2014){Strugarek}, {Brun}, {Matt}, \&
  {R{\'e}ville}}]{Strugarek2}
{Strugarek}, A., {Brun}, A.~S., {Matt}, S.~P., \& {R{\'e}ville}, V. 2014, apj,
  795, 86

\bibitem[{{Strugarek} {et~al.}(2015){Strugarek}, {Brun}, {Matt}, \&
  {R{\'e}ville}}]{Strugarek}
{Strugarek}, A., {Brun}, A.~S., {Matt}, S.~P., \& {R{\'e}ville}, V. 2015, apj,
  815, 111

\bibitem[{{Suzuki, T.K.}(2013)}]{Suzuki}
{Suzuki, T.K.} 2013, Astronomische Nachrichten, 334, 81

\bibitem[{{Tarter, J. C.} {et~al.}(2007){Tarter, J. C.}, {Backus, P. R.},
  {Mancinelli, R. L.}, {Aurnou, J. M.}, {Backman, D. E.}, {Basri, G. S.},
  {Boss, A. P.}, {Clarke, A.}, {Deming, D.}, {Doyle, L. R.}, {Feigelson, E.
  D.}, {Freund, F.}, {Grinspoon, D. H.}, {Haberle, R. M.}, {Hauck, S. A.},
  {Heath, M. J.}, {Henry, T. J.}, {Hollingsworth, J. L.}, {Joshi, M. M.},
  {Kilston, S.}, {Liu, M. C.}, {Meikle, E.}, {Reid, I. N.}, {Rothschild, L.
  J.}, {Scalo, J.}, {Segura, A.}, {Tang, C. M.}, {Tiedje, J. M.}, {Turnbull, M.
  C.}, {Walkowicz, L. M.}, {Weber, A. L.}, \& {Young, R. E.}}]{Tarter}
{Tarter, J. C.}, {Backus, P. R.}, {Mancinelli, R. L.}, {et~al.} 2007,
  Astrobiology, 7, 30

\bibitem[{{Thomsen, M. F.}(2004)}]{Thomsen}
{Thomsen, M. F.} 2004, Space Weather, 2, S11004

\bibitem[{{Turner, J. D.} {et~al.}(2021){Turner, J. D.}, {Zarka, P.},
  {Griessmeier, J. M.}, {Lazio, J.}, {Cecconi, B.}, {Emilio Enriquez, J.},
  {Girard, J. N.}, {Jayawardhana, R.}, {Lamy, L.}, {Nichols, J. D.}, \& {de
  Pater, I.}}]{Turner3}
{Turner, J. D.}, {Zarka, P.}, {Griessmeier, J. M.}, {et~al.} 2021, A\&A, 645,
  A59

\bibitem[{{van Haarlem, M. P.} {et~al.}(2013){van Haarlem, M. P.}, {Wise, M.
  W.}, {Gunst, A. W.}, {Heald, G.}, {McKean, J. P.}, {Hessels, J. W. T.}, {de
  Bruyn, A. G.}, {Nijboer, R.}, {Swinbank, J.}, {Fallows, R.}, {Brentjens, M.},
  {Nelles, A.}, {Beck, R.}, {Falcke, H.}, {Fender, R.}, {Hörandel, J.},
  {Koopmans, L. V. E.}, {Mann, G.}, {Miley, G.}, {Röttgering, H.}, {Stappers,
  B. W.}, {Wijers, R. A. M. J.}, {Zaroubi, S.}, {van den Akker, M.}, {Alexov,
  A.}, {Anderson, J.}, {Anderson, K.}, {van Ardenne, A.}, {Arts, M.}, {Asgekar,
  A.}, {Avruch, I. M.}, {Batejat, F.}, {Bähren, L.}, {Bell, M. E.}, {Bell, M.
  R.}, {van Bemmel, I.}, {Bennema, P.}, {Bentum, M. J.}, {Bernardi, G.}, {Best,
  P.}, {Bîrzan, L.}, {Bonafede, A.}, {Boonstra, A.-J.}, {Braun, R.}, {Bregman,
  J.}, {Breitling, F.}, {van de Brink, R. H.}, {Broderick, J.}, {Broekema, P.
  C.}, {Brouw, W. N.}, {Brüggen, M.}, {Butcher, H. R.}, {van Cappellen, W.},
  {Ciardi, B.}, {Coenen, T.}, {Conway, J.}, {Coolen, A.}, {Corstanje, A.},
  {Damstra, S.}, {Davies, O.}, {Deller, A. T.}, {Dettmar, R.-J.}, {van Diepen,
  G.}, {Dijkstra, K.}, {Donker, P.}, {Doorduin, A.}, {Dromer, J.}, {Drost, M.},
  {van Duin, A.}, {Eislöffel, J.}, {van Enst, J.}, {Ferrari, C.}, {Frieswijk,
  W.}, {Gankema, H.}, {Garrett, M. A.}, {de Gasperin, F.}, {Gerbers, M.}, {de
  Geus, E.}, {Grießmeier, J.-M.}, {Grit, T.}, {Gruppen, P.}, {Hamaker, J. P.},
  {Hassall, T.}, {Hoeft, M.}, {Holties, H. A.}, {Horneffer, A.}, {van der
  Horst, A.}, {van Houwelingen, A.}, {Huijgen, A.}, {Iacobelli, M.}, {Intema,
  H.}, {Jackson, N.}, {Jelic, V.}, {de Jong, A.}, {Juette, E.}, {Kant, D.},
  {Karastergiou, A.}, {Koers, A.}, {Kollen, H.}, {Kondratiev, V. I.},
  {Kooistra, E.}, {Koopman, Y.}, {Koster, A.}, {Kuniyoshi, M.}, {Kramer, M.},
  {Kuper, G.}, {Lambropoulos, P.}, {Law, C.}, {van Leeuwen, J.}, {Lemaitre,
  J.}, {Loose, M.}, {Maat, P.}, {Macario, G.}, {Markoff, S.}, {Masters, J.},
  {McFadden, R. A.}, {McKay-Bukowski, D.}, {Meijering, H.}, {Meulman, H.},
  {Mevius, M.}, {Middelberg, E.}, {Millenaar, R.}, {Miller-Jones, J. C. A.},
  {Mohan, R. N.}, {Mol, J. D.}, {Morawietz, J.}, {Morganti, R.}, {Mulcahy, D.
  D.}, {Mulder, E.}, {Munk, H.}, {Nieuwenhuis, L.}, {van Nieuwpoort, R.},
  {Noordam, J. E.}, {Norden, M.}, {Noutsos, A.}, {Offringa, A. R.}, {Olofsson,
  H.}, {Omar, A.}, {Orrú, E.}, {Overeem, R.}, {Paas, H.}, {Pandey-Pommier,
  M.}, {Pandey, V. N.}, {Pizzo, R.}, {Polatidis, A.}, {Rafferty, D.},
  {Rawlings, S.}, {Reich, W.}, {de Reijer, J.-P.}, {Reitsma, J.}, {Renting, G.
  A.}, {Riemers, P.}, {Rol, E.}, {Romein, J. W.}, {Roosjen, J.}, {Ruiter, M.},
  {Scaife, A.}, {van der Schaaf, K.}, {Scheers, B.}, {Schellart, P.},
  {Schoenmakers, A.}, {Schoonderbeek, G.}, {Serylak, M.}, {Shulevski, A.},
  {Sluman, J.}, {Smirnov, O.}, {Sobey, C.}, {Spreeuw, H.}, {Steinmetz, M.},
  {Sterks, C. G. M.}, {Stiepel, H.-J.}, {Stuurwold, K.}, {Tagger, M.}, {Tang,
  Y.}, {Tasse, C.}, {Thomas, I.}, {Thoudam, S.}, {Toribio, M. C.}, {van der
  Tol, B.}, {Usov, O.}, {van Veelen, M.}, {van der Veen, A.-J.}, {ter Veen,
  S.}, {Verbiest, J. P. W.}, {Vermeulen, R.}, {Vermaas, N.}, {Vocks, C.},
  {Vogt, C.}, {de Vos, M.}, {van der Wal, E.}, {van Weeren, R.}, {Weggemans,
  H.}, {Weltevrede, P.}, {White, S.}, {Wijnholds, S. J.}, {Wilhelmsson, T.},
  {Wucknitz, O.}, {Yatawatta, S.}, {Zarka, P.}, {Zensus, A.}, \& {van Zwieten,
  J.}}]{Haarlem}
{van Haarlem, M. P.}, {Wise, M. W.}, {Gunst, A. W.}, {et~al.} 2013, A\&A, 556,
  A2

\bibitem[{{Varela} {et~al.}(2015){Varela}, {Pantellini}, \&
  {Moncuquet}}]{Varela}
{Varela}, J., {Pantellini}, F., \& {Moncuquet}, M. 2015, Planet. Space Sci.,
  119, 264

\bibitem[{{Varela} {et~al.}(2016{\natexlab{a}}){Varela}, {Pantellini}, \&
  {Moncuquet}}]{Varela4}
{Varela}, J., {Pantellini}, F., \& {Moncuquet}, M. 2016{\natexlab{a}}, Planet.
  Space Sci., 129, 74

\bibitem[{{Varela} {et~al.}(2016{\natexlab{b}}){Varela}, {Pantellini}, \&
  {Moncuquet}}]{Varela2}
{Varela}, J., {Pantellini}, F., \& {Moncuquet}, M. 2016{\natexlab{b}}, Planet.
  Space Sci., 120, 78

\bibitem[{{Varela} {et~al.}(2016{\natexlab{c}}){Varela}, {Pantellini}, \&
  {Moncuquet}}]{Varela3}
{Varela}, J., {Pantellini}, F., \& {Moncuquet}, M. 2016{\natexlab{c}}, Planet.
  Space Sci., 122, 46

\bibitem[{{Varela} {et~al.}(2016{\natexlab{d}}){Varela}, {Reville}, {Brun.},
  {Pantellini}, \& {Zarka}}]{Varela5}
{Varela}, J., {Reville}, V., {Brun.}, A.~S., {Pantellini}, F., \& {Zarka}, P.
  2016{\natexlab{d}}, A\&A, 595, A69

\bibitem[{{Varela, J.} {et~al.}(2018){Varela, J.}, {Reville, V.}, {Brun, A.
  S.}, {Zarka, P.}, \& {Pantellini, F.}}]{Varela6}
{Varela, J.}, {Reville, V.}, {Brun, A. S.}, {Zarka, P.}, \& {Pantellini, F.}
  2018, A\& A, 616, A182

\bibitem[{{Varela, J.} {et~al.}(2022){Varela, J.}, {Reville, V.}, {Brun, A.
  S.}, {Zarka, P.}, \& {Pantellini, F.}}]{Varela7}
{Varela, J.}, {Reville, V.}, {Brun, A. S.}, {Zarka, P.}, \& {Pantellini, F.}
  2022, A\& A, 659, A10

\bibitem[{{Vedantham, H. K.} {et~al.}(2020){Vedantham, H. K.}, {Callingham, J.
  R.}, {Shimwell, T. W.}, {Tasse, C.}, {Pope, B. J. S.}, {Bedell, M.},
  {Snellen, T.}, {Best, P.}, {Hardcastle, M. J.}, {Haverkorn, M.}, {Mechev,
  A.}, {OSullivan, S. P.}, {Rottgering, H. J. A.}, \& {White, G.
  J.}}]{Vedantham}
{Vedantham, H. K.}, {Callingham, J. R.}, {Shimwell, T. W.}, {et~al.} 2020,
  Nature Astronomy, 4, 577

\bibitem[{{Vidotto} {et~al.}(2012){Vidotto}, {Fares}, {Jardine}, {Donati},
  {Opher}, {Moutou}, {Catala}, \& {Gombosi}}]{Vidotto6}
{Vidotto}, A.~A., {Fares}, R., {Jardine}, M., {et~al.} 2012, MNRAS, 423, 3285

\bibitem[{{Vidotto, A. A.} {et~al.}(2013){Vidotto, A. A.}, {Jardine, M.},
  {Morin, J.}, {Donati, J. F.}, {Opher, M.}, \& {Gombosi, T. I.}}]{Vidotto9}
{Vidotto, A. A.}, {Jardine, M.}, {Morin, J.}, {et~al.} 2013, Monthly Notices of
  the Royal Astronomical Society, 438, 1162

\bibitem[{{Wang, Y. L.} {et~al.}(2003){Wang, Y. L.}, {Raeder, J.}, {Russell, C.
  T.}, {Phan, T. D.}, \& {Manapat, M.}}]{Wang6}
{Wang, Y. L.}, {Raeder, J.}, {Russell, C. T.}, {Phan, T. D.}, \& {Manapat, M.}
  2003, Journal of Geophysical Research: Space Physics, 108, SMP 8

\bibitem[{{Wang, Y. M.} {et~al.}(2003){Wang, Y. M.}, {Ye, P. Z.}, {Wang, S.},
  \& {Xue, X. H.}}]{Wang4}
{Wang, Y. M.}, {Ye, P. Z.}, {Wang, S.}, \& {Xue, X. H.} 2003, Geophysical
  Research Letters, 30, 1

\bibitem[{{Watanabe, K.} \& {Sato, T.}(1990)}]{Watanabe}
{Watanabe, K.} \& {Sato, T.} 1990, Journal of Geophysical Research: Space
  Physics, 95, 75

\bibitem[{{Wu}(1979)}]{Wu5}
{Wu}, C.~S. 1979, 230, 621

\bibitem[{{Wu, C.} \& {Lepping, R. P.}(2015)}]{Wu}
{Wu, C.} \& {Lepping, R. P.} 2015, Sol Phys, 290, 1243

\bibitem[{{Yang} {et~al.}(2013){Yang}, {Cowan}, \& {Abbot}}]{Yang}
{Yang}, J., {Cowan}, N.~B., \& {Abbot}, D.~S. 2013, The Astrophysical Journal,
  771, L45

\bibitem[{{Zarka}(1998)}]{Zarka5}
{Zarka}, P. 1998, J. Geophys. Res., 103, 20159

\bibitem[{{Zarka}(2007)}]{Zarka9}
{Zarka}, P. 2007, Planetary and Space Science, 55, 598

\bibitem[{{Zarka}(2018)}]{Zarka8}
{Zarka}, P. 2018, Handbook of Exoplanets, 1775

\bibitem[{{Zarka} {et~al.}(2020){Zarka}, {Denis}, {Tagger}, {Girard}, {Coffre},
  {Dumez-Viou}, {Taffoureau}, {Charrier}, {Bondonneau}, {Briand}, {Casoli},
  {Cecconi}, {Cognard}, \& {Corbel}}]{Zarka11}
{Zarka}, P., {Denis}, L., {Tagger}, M., {et~al.} 2020, New Telescopes on the
  Frontier, Session J01, URSI GASS, Rome,
  http://www.ursi.org/proceedings/procGA20/papers/URSIGASS2020SummaryPaperNenuFARnew.pdf

\bibitem[{{Zarka} {et~al.}(2015){Zarka}, {Lazio}, \& {Hallinan}}]{Zarka10}
{Zarka}, P., {Lazio}, T. J.~W., \& {Hallinan}, G. 2015, Proceedings of
  Advancing Astrophysics with the Square Kilometre Array (AASKA14), Giardini
  Naxos, Italy, https://pos.sissa.it/215/120/pdf, id.120

\bibitem[{{Zarka} {et~al.}(2001){Zarka}, {Treumann}, {Ryabov}, \&
  {Ryabov}}]{Zarka3}
{Zarka}, P., {Treumann}, R.~A., {Ryabov}, B.~P., \& {Ryabov}, V.~B. 2001,
  Astrophys. Space Sci., 277, 293

\end{thebibliography}

\begin{appendix}
\section{Numerical model validation}

\begin{figure}[h]
\centering
\resizebox{\hsize}{!}{\includegraphics[width=\columnwidth]{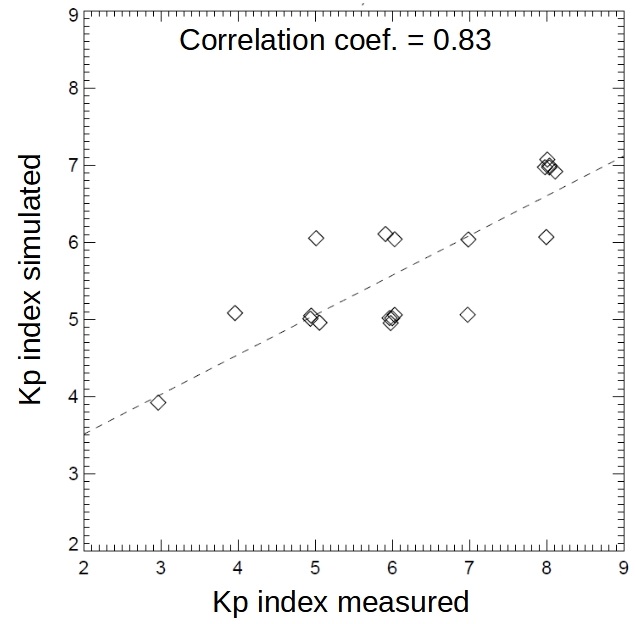}}
\caption{Correlation between the $K_{p}$ index obtained in the simulations with respect to the measured values.} 
\label{x}
\end{figure}

The numerical model used in this study was also applied in the analysis of the interaction between the solar wind and the Earth magnetosphere \citep{Varela7}. Part of \citet{Varela7} study was dedicated to analyze the perturbation induced in the magnetosphere by several CMEs that impacted the Earth from $1997$ to $2020$. The simulations results were compared with observational data to validate the numerical model, in particular the $K_{p}$ index. The $K_{p}$ index provides the global geomagnetic activity taking values from $0$ if the geomagnetic activity is weak to $9$ if the geomagnetic activity is extreme \citep{Menvielle,Thomsen}. The $K_{p}$ index was calculated in the simulations as the lowest latitude with open magnetic field lines in the Earth surface at the North Hemisphere. Figure \ref{x} shows the correlation between the $K_{p}$ index obtained in the simulations with respect to the measured values. The statistical analysis finds a correlation coefficient of $0.83$, that is to say, a reasonable agreement between simulations and observational data. Consequently, the numerical model is valid to reproduce the global structures of the Earth magnetosphere during extreme space weather conditions, also suitable to analyze the interaction of the stellar wind with exoplanet magnetospheres if the intrinsic magnetic field is similar to the Earth. 

\section{Calculation of the magnetopause standoff distance}

The theoretical approximation of the magnetopause standoff distance is calculated as the balance between the dynamic pressure of the SW ($P_{d} = m_{p} n_{sw}  v_{sw}^{2}/2$), the thermal pressure of the SW ($P_{th,sw} = m_{p} n_{sw}  v_{th,sw}^{2}/2 = m_{p} n_{sw} c_{sw}^{2}/\gamma$), and the magnetic pressure of the IMF ($P_{mag,sw} = B_{sw}^{2}/(2 \mu_{0})$ with respect to the magnetic pressure of a dipolar magnetic field ($P_{mag,ex} = \alpha \mu_{0} M_{ex}^{2} / 8 \pi^2 r^{6} $) and the thermal pressure of the magnetosphere ($P_{th,MSP} = m_{p} n_{MSP}  v_{th,MSP}^{2}/2$). This results in the expression:
\begin{equation}
\label{eqn:pressure}
P_{d} + P_{mag,sw} + P_{th,sw} = P_{mag,ex} + P_{th,MSP}
\end{equation}
\begin{equation}
\label{eqn:balance}
\frac{R_{mp}}{R_{ex}} = \left[ \frac{\alpha \mu_{0} M_{ex}^{2}}{4 \pi^2 \left( m_{p} n_{sw}  v_{sw}^{2} + \frac{B_{sw}^{2}}{\mu_{0}} + \frac{2 m_{p} n_{sw}  c_{sw}^{2}}{\gamma} - m_{p} n_{BS} v_{th,MSP}^{2} \right)} \right]^{(1/6)}
\end{equation}
with $M_{ex}$ the exoplanet dipole magnetic field moment, $r = R_{mp} / R_{ex}$ , and $\alpha$ the dipole compression coefficient ($\alpha \approx 2$ \citep{Gombosi}). This approximation does not include the effect of the reconnections between the IMF with the exoplanet magnetic fields, thus the expression assumes a compressed dipolar magnetic field, ignoring the orientation of the IMF. Here, the approximation is only valid if the IMF intensity is rather low and the magnetopause standoff distance should be calculated using simulations for extreme space weather conditions.

The magnetopause standoff distance is defined in the simulations analysis as the last close magnetic field line on the exoplanet dayside at $0^{o}$ longitude in the ecliptic plane.

\end{appendix}


\end{document}